\documentclass{article}
\usepackage[utf8]{inputenc}
\usepackage{amsmath,amssymb,amsthm}
\usepackage{latexsym}            

\usepackage{epsfig}
\usepackage{epstopdf}
\usepackage{amsfonts,amscd, makecell}
\usepackage[ruled,vlined]{algorithm2e}
\usepackage{color,comment}
\usepackage{amsbsy,bm}
\usepackage{subfigure}
\usepackage[utf8]{inputenc}          
\usepackage[T1]{fontenc}  

\usepackage{xcolor}
\usepackage{overpic}
\newcommand\dd{\mathrm{d}}
\newcommand\pp{\partial}

\newcommand\x{\bm{x}}

\newcommand\uvec{\mathbf{u}}
\newcommand\X{\mathbf{X}}
\newcommand\y{\bm{y}}
\newcommand\qvec{\bm{q}}
\newcommand\rvec{\bm{r}}
\newcommand\Qvec{\bm{Q}}

\newcommand\xvec{\bm{x}}
\newcommand\Xvec{\bm{X}}

\newcommand\Vvec{\mathbf{V}}

\def\theequation{\arabic{equation}}

\newcommand{\be}{\begin{eqnarray}}
\newcommand{\ee}{\end{eqnarray}}
\newcommand{\ben}{\begin{eqnarray*}}
\newcommand{\een}{\end{eqnarray*}}

\newtheorem{remark}{Remark}[section]

 \allowdisplaybreaks \allowdisplaybreaks[4]
 \usepackage{authblk}

\providecommand{\keywords}[1]
{
  \small	
  \textbf{\textit{Keywords---}} #1
}

\title{Micro-Macro Modeling of Polymeric Fluids and Shear-Induced Microscopic Behaviors}
\author[1,2]{Xuelian Bao}
\author[1,2,3,4,5]{Huaxiong Huang}
\author[6]{Zilong Song}
\author[7]{Shixin Xu*}
\affil[1]{School of Mathematical Sciences, Beijing Normal University, Beijing 100875, China}
\affil[2]{ Research Centre for Mathematics, Beijing
Normal University, Zhuhai, China}
\affil[3]{Guandong Key Lab for Data Science and Technology, BNU-HKBU United International College, Zhuhai, China }
\affil[4]{Laboratory of Mathematics and Complex Systems, MOE, Beijing Normal University, 100875 Beijing, China}
\affil[5]{Department of
Mathematics and Statistics York University, Toronto, ON, M3J 1P3, Canada}
\affil[6]{Mathematics and Statistics Department, Utah State University, Old Main Hill Logan, UT 84322 }
\affil[7]{Zu Chongzhi Center for Mathematics and Computational Sciences (CMCS), Duke Kunshan University, 8 Duke Ave., Kunshan, Jiangsu, 215316, China}

\date{ }

\begin{document}
 
\maketitle
\begin{abstract}
 
This article delves into the micro-macro modeling of polymeric fluids, considering various microscopic potential energies, including the classical Hookean potential, as well as newly proposed modified Morse and Elastic-plastic potentials. These proposed potentials encompass microscopic-scale bond-breaking processes.
The development of a thermodynamically consistent micro-macro model is revisited, employing the energy variational method. To validate the model's predictions, we conduct numerical simulations utilizing a deterministic particle-FEM method.
Our numerical findings shed light on the distinct behaviors exhibited by polymer chains at the micro-scale in comparison to the macro-scale velocity and induced shear stresses of fluids under shear flow. Notably, we observe that polymer elongation, rotation, and bond breaking contribute to the zero polymer-induced stress in the micro-macro model when employing Morse and Elastic-plastic potentials. Furthermore, at high shear rates, polymer rotation is found to induce shear-thinning behavior in the model employing the classical Hookean potential.

 
\end{abstract}

 \section{Introduction}

Models for complex fluids can be categorized into pure macroscopic models \cite{Keunings1997,Lielens1998,Sizaire1999} and micro-macro models \cite{bird1992transport, le2012micro}.  
This article focuses on micro-macro models of complex fluids, with one of the simplest models being the micro-macro model for dilute polymeric fluids using a bead-spring chain representation of polymer molecules \cite{lin-liu-zhang}.
More precisely, at the microscopic level, the polymer molecule is represented as a linear chain of $N$ beads, which correspond to molecular segments consisting of multiple monomers \cite{ottinger1996}, interconnected by $N-1$ entropic springs.
The molecular configuration is characterized by end-to-end vectors $\qvec_i =\rvec_{i+1}-\rvec_i$ $(i = 1, 2, \cdots, N-1)$ where $\rvec_i$ denotes the position vector of the $i$-th bead.
The micro-macro model is given by \cite{bird1987}
\begin{equation}\label{CPsystemwithf_1}
\left\{
\begin{aligned}
&\rho(\uvec_t+\uvec \cdot \nabla_{\x} \uvec)+\nabla_{\x} p= \eta_s \Delta_{\x} \uvec+\nabla_{\x} \cdot {\bm \tau}_p,\\
& \nabla_{\x}\cdot\uvec=0,\\
&{\bm \tau}_p = \sum_{i=1}^{N-1}  \lambda_p \mathbb{E} (\nabla_{\qvec_i} \Psi \otimes \qvec_i )  = \sum_{i=1}^{N-1} \lambda_p \int_{\mathbb{R}^d} \cdots \int_{\mathbb{R}^d} f \nabla_{\qvec_i} \Psi\otimes \qvec_i \dd \qvec_1 \cdots \dd \qvec_{N-1},\\
&f_t + \nabla_{\x} f \cdot \uvec + \sum_{i=1}^{N-1} \nabla_{\qvec_i} \cdot ( (\nabla_{\x} \uvec)^T \cdot  \qvec_i f) =  \frac{1}{\zeta}  \sum_{i=1}^{N-1} \nabla_{\qvec_i} \cdot \bigg(  \sum_{k=1}^{N-1}   A_{ik}  [k_B T \nabla_{\qvec_k} f+ f \nabla_{\qvec_k} \Psi] \bigg).
\end{aligned}
\right.
\end{equation}
The first two equations are incompressible Navier-Stokes equations for the  macroscopic motion of the fluid. $\uvec$ denotes the fluid velocity field, $p$ is the pressure, $\rho$ is the density of the fluid. The elastic stress ${\bm \tau}_p$, depending on the microscopic configuration of polymer chains, is described by the third equation. The dynamics of the polymer number density distribution $f$ is  modeled by a  Fokker-Planck equation  with a drift term depending on the macroscopic velocity field $\uvec$. 
Moreover, $\lambda_p$ is a parameter related to the polymer density, 
$\zeta$ is a constant related to the polymer relaxation time,  $k_B$ is the Boltzmann constant, $T$ is the absolute temperature and $A_{ik}$ are the elements of the Rouse matrix defined by
$$
A_{ij} = \left\{
\begin{aligned}
& 2 \qquad \quad \mbox{if} \ i=j,\\
& -1 \qquad \mbox{if} \ i=j \pm 1, \\
& 0 \qquad \quad \mbox{otherwise.} 
\end{aligned}
\right.
$$

This micro-macro model assumes that the distribution of configurations is independent of the location of the polymer chains in space, thus, the polymer number density distribution $f = f(\x,\qvec_1, \cdots, \qvec_{N-1}, t)$ and the spring potential energy $\Psi = \Psi(\qvec_1, \cdots, \qvec_{N-1})$ only depend on the connector vectors $\qvec_1, \cdots, \qvec_{N-1}$. 
And in the absence of external forces, it also assumes that the center of mass of the polymer chain moves with the solution velocity at each $\x$.  

Alternatively, the microscopic dynamics can also be described by a stochastic differential equation (SDE), or Langevin dynamics, given by \cite{Bris2012, xu2016multi}
\begin{equation}\label{SDE}
\begin{aligned}
\dd \qvec_i(\x, t) & = ( - \uvec \cdot \nabla_{\x} \qvec_i(\x, t) + (\nabla_{\x} \uvec)^T \cdot \qvec_i(\x, t) + 2 \zeta^{-1}   \tilde{\bm F}_i) \dd t + \sqrt{4 k_B T \zeta^{-1}} \dd {\bf W}_{it},
\end{aligned}
\end{equation}
for $i=1, \cdots, N-1$. Here, $\dd {\bf W}_{it}$ is the standard multidimensional white noise and the entropic forces are given by 
$$
 \tilde{\bm F}_i =  \left\{
\begin{aligned}
& -2\bm F_i + \bm F_{i+1}, \qquad \qquad \ i=1,\\
&\bm F_{i-1}- 2  \bm F_i + \bm F_{i+1}, \qquad \ i=2, \cdots, N-2, \\
& \bm F_{i-1}- 2  \bm F_i,  \qquad \qquad \quad  i=N-1,
\end{aligned}
\right.
$$
where $\bm F_i = \nabla_{\qvec_i} \Psi$ is the entropic force for the $i-$th segment.


In this article, we first rederive the thermodynamically consistent micro-macro model \eqref{CPsystemwithf_1} with two beads for dilute polymeric fluids via the energy variational method \cite{shen2020energy, shen2022energy}. 
This energy variational method follows a similar approach proposed by H.C. \"{O}ttinger and M. Grmela \cite{Grmela1997I, Grmela1997II}. 
Precisely, it begins with defining two functionals for the total energy and dissipation of the system and introducing the kinematic equations based on physical laws of conservation. The specific forms of the flux and stress functions in the kinematic equations are obtained by taking the time derivative of the total energy functional and comparing it with the defined dissipation functional. More details of this method can be found in Ref. \cite{shen2020energy}. Our energy variational method consistently yields the same micro-macro system obtained by the energetic variational approach \cite{wang2020jcp, wang2021two, bao2022}.

Laboratory experiments and numerical simulations demonstrate the significant influence of the molecular model at the microscopic scale on the properties of macroscopic polymeric fluids. Specifically, the classical molecular model in the micro-macro model \eqref{CPsystemwithf_1} is the bead-spring model with various microscopic potential energies that describe the interactions between bead pairs.
The classical microscopic potential corresponds to the Hookean potential, given by
$$\Psi = \sum_{i=1}^{N-1} \frac{H}{2} | \qvec_i |^2,$$
which leads to the macroscopic viscoelastic model known as the Oldroyd-B model \cite{Laso1993, ottinger1996, ACR2006, Ammar2010}, where $H > 0$ represents the elastic constant. However, the Hookean potential exhibits an undesired infinite elongation problem and is only applicable for small deformations from the equilibrium state.
Another fully investigated nonlinear potential is the Finite Extensible Nonlinear Elastic (FENE) potential, which accounts for the finite extensibility of polymer molecules using the positive parameter $Q_0$:
$$\Psi = \sum_{i=1}^{N-1} -\frac{H Q_0^2}{2} \ln \left(1- \left(\frac{|\qvec_i|}{Q_0}\right)^2\right).$$
From a physical perspective, the FENE potential is more realistic but introduces an additional singular nonlinearity \cite{Doyle1998, Lielens1998, Sizaire1999}. Moreover, under conditions where polymer chains are elongated far from the equilibrium state, irreversible bond breaking of polymers may occur under shear stress, such as the dissociation of blood clots \cite{nguyen2017single, xu2017model}. In such scenarios, the pure elastic (Hookean) and finite extensible (FENE) models are no longer suitable.
However, the microscopic mechanisms underlying the mechanical failure of soft polymer materials remain poorly understood. This is primarily due to the lack of experimental tools capable of preparing well-controlled model systems and observing the failure process in real-time at the microscopic scale, particularly in the field of biology \cite{needleman2017active}.

There are various nonlinear intermolecular potentials available to describe long-range interactions in molecular dynamics. Among them, the Morse potential has been extensively utilized in quantum mechanics since its initial application in solving the Schr\"{o}dinger equation. It has also been successfully employed to model hydrogen bonds in DNA chains, specifically in the study of base connections \cite{Filho2013, Theodorakopoulos2000}. The Morse potential, a classical nonlinear potential energy for diatomic molecules, explicitly incorporates the effects of bond breaking, including the existence of unbound states. While the Morse potential provides a smooth description of bond breaking, there are cases where a more explicit representation of the abrupt breaking process is required. In such situations, a modified Morse potential energy is proposed. In another case, assuming that the internal interactions between two beads remain constant when the end-to-end distance between them exceeds a critical value, we observe a behavior analogous to that of visco-plastic fluids \cite{Papanastasiou1997}. Visco-plastic fluids exhibit solid-like behavior at low stresses and fluid-like response at high stresses. This behavior is commonly observed in industries such as petroleum and chemical processing \cite{viscoplastic, Glowinski2011}. In line with the well-known visco-elastic and visco-plastic models, we refer to this potential energy as the Elastic-plastic potential.

The second objective of this article is to compare the behaviors of polymer chains with different microscopic potential energies, as well as the resulting velocity and induced shear stress of fluids at the macroscopic scale under shear flow, through numerical simulations. In order to investigate polymer chain configurations far from equilibrium and the effects of irreversible bond breaking, we select a modified Morse potential and the Elastic-plastic potential as the potential energies in this study. Additionally, we examine the behaviors of polymer chains, velocity, and shear stress using the classical Hookean potential for comparison.

The rest of this article is organized as follows. We present the rederivation of 
the thermodynamically consistent micro-macro model \eqref{CPsystemwithf_1} for dilute polymeric fluids by employing the energy variational method and its nondimensionalization in Section 2.
Then we summarize the numerical method in Section 3.  Results and discussion are presented in Section 4, including 
numerical results of the micro-macro model with Morse, Elastic-plastic, and Hookean potentials under simple shear flow with constant and periodic shear rate. 
And the effects of microscopic bond breaking on the velocity and shear stresses of the macroscopic fluid are illustrated by investigating the configurations of polymer chains.
Finally, the concluding remarks are given in Section 5.

\section{Mathematical Model}

In the current work, we focus on the simple dumbbell model, where a polymer chain is modeled by a dumbbell on the microscopic scale, consisting of two beads connected by a single spring, namely, the micro-macro model \eqref{CPsystemwithf_1} with $N=2$. Since there is only a single $\qvec_1$ in this case, we write $\qvec_1$ as $\qvec$  for simplicity.

\subsection{Assumptions and laws of conservation}\label{assulaw}

We start from the kinematic assumptions on the macroscopic scale. In domain $\Omega$, we have the laws of   conservation:
\begin{equation}
    \begin{aligned}
    & \rho\frac{D \uvec}{Dt} = \nabla_{\x}\cdot \bm\tau = \nabla_{\x}\cdot \bm\tau_s + \nabla_{\x}\cdot \bm\tau_p, \\
    & \nabla_{\x} \cdot \uvec = 0.
    \end{aligned}
\end{equation}
where $\rho$ is the density, $\uvec$ is the fluid velocity, and $\frac{D \uvec}{Dt} = \frac{\partial \uvec}{\partial t} + \uvec \cdot \nabla_{\x} \uvec$ is the material derivative. This is the conservation of momentum with stress tensor $\bm \tau$, which consists of contributions from the macroscopic viscosity fluid $\bm \tau_s$ and the microscopic polymer $\bm \tau_p$. 
We set the no-slip boundary condition for $\uvec$, i.e. $\uvec|_{\partial \Omega} = 0$.

On the microscopic scale, assume that the number density of polymer chains is spatially homogeneous, and thus, let $n(\x, \qvec, t)$ be the number of polymers which have end-to-end vector $\qvec=\rvec_2-\rvec_1$ at position $\x \in \Omega$ at time $t$. Then for any $\x \in \Omega$ in the physical space, we have
\begin{equation}\label{int_n}
     \int_{\mathbb{R}^d} n(\x, \qvec, t) \dd \qvec= N_q,
\end{equation}
where $N_q$ is the total number of all polymers. 
Denote $f = f(\x,\qvec, t)$ as the number density distribution function, namely, $f(\x, \qvec, t) = n(\x, \qvec, t)/N_q$.

Borrowing the flow map definitions from the energetic variational approach \cite{wang2021two, wang2020jcp, wangEVI}, let $\xvec(\Xvec, t)$ be the flow
map at the physical space and $\qvec(\Xvec, \Qvec, t)$ be the flow map at the configurational space,  where $\Xvec$ and $\Qvec$ are Lagrangian coordinates in physical and configurational space respectively. 
Hence, we have
$$
f(\x, \qvec, t) = f(\x(\X, t), \qvec(\x(\X, t), \Qvec, t), t).
$$
Given $\x \in \Omega$ and $\x(\X, t=0)=\X$. 
Assume arbitrary unit volume $\Omega_{\Qvec}$ in the Lagrangian coordinate of the configurational space, we have 
\begin{equation}\label{f0}
    \begin{aligned}
        \int_{\Omega_{\Qvec}} f_0(\X, \Qvec) \dd\Qvec & =  \int_{\Omega_{\qvec}} f(\x, \qvec, t) \dd\qvec = \int_{\Omega_{\Qvec}} f \det G \ \dd\Qvec,
    \end{aligned}
\end{equation}
where $f_0(\X, \Qvec) = f(\X, \Qvec, 0)$  is the initial number distribution function, $\Omega_{\qvec}$ is the corresponding volume of $\Omega_{\Qvec}$ at time $t$ and the Jacobi matrix $G = \nabla_{\Qvec} \qvec$. 

Since $\Omega_{\Qvec}$ and $f_0(\X, \Qvec)$ are independent of time $t$, taking derivative of \eqref{f0}with respect to $t$, we have
\begin{equation}\label{dtf0}
    \begin{aligned}
        0 & = \frac{d}{dt} \int_{\Omega_{\Qvec}}  f \det G \ \dd\Qvec = \int_{\Omega_{\Qvec}} \frac{d}{dt}(f \det G) \ \dd\Qvec\\
        & = \int_{\Omega_{\Qvec}} \left(f_t +\uvec\cdot \nabla_{\x} f+\Vvec\cdot\nabla_{\qvec} f +G^{-T}:\frac{dG}{dt}f  \right)\det G \ \dd\Qvec\\
        & = \int_{\Omega_{\Qvec}} \left(f_t +\uvec\cdot \nabla_{\x} f+\Vvec\cdot\nabla_{\qvec} f +f\nabla_{\qvec}\cdot\Vvec\right)\det G \dd \Qvec \\
        & = \int_{\Omega_{\qvec}} \left(f_t +\uvec\cdot \nabla_{\x} f+\Vvec\cdot\nabla_{\qvec} f +f\nabla_{\qvec}\cdot\Vvec\right) \dd \qvec\\
      & = \int_{\Omega_{\qvec}} \left(f_t + \nabla_{\x} \cdot ( f\uvec)  +\nabla_{\qvec}\cdot (f \Vvec)\right) \dd \qvec\\
    \end{aligned}
\end{equation}
where  the incompressibility condition $\nabla_{\x} \cdot \uvec = 0$ is used in the last equality and $\Vvec$ 
is the microvelocity.

Without dissipation, on the microscopic scale, the macroscopic induced velocity is obtained via the Cauchy-Born relation \cite{tm11}, which assumes that the microscopic vector $\Qvec$ is
a deformed vector by the macroscopic motion:
$$\qvec = F\Qvec,$$
where $F$ is the deformation gradient tensor and $F = \nabla_{\X} \x$. 
Then the macroscopic induced velocity yields that
$$\frac{d}{dt} (F \Qvec) = (\frac{d}{dt} F) \Qvec =  (\nabla_{\x} \uvec)^T \cdot \qvec.$$ 
 However, when it comes to the micro-macro coupling, for viscoelastic flows, the microvelocity is not equal to the macroscopic induced velocity due to the dissipation induced by the relative friction of microscopic beads to the macroscopic flow. 
Hence, denote $\tilde{\Vvec}$ be the difference between the microvelocity and the macroscopic induced velocity as follows,
\begin{equation}\label{macroinduced}
 \tilde{\Vvec} = \Vvec-(\nabla_{\x} \uvec)^T \cdot \qvec.
\end{equation}
Then by \eqref{dtf0} and \eqref{macroinduced}, due to the arbitrariness of $\Omega_{\Qvec}$, the law of conservation  on the microscopic scale yields
\begin{equation}\label{conservationf}
  f_t + \nabla_{\x} \cdot ( f\uvec)  + \nabla_{\qvec} \cdot ( (\nabla_{\x} \uvec)^T \cdot \qvec f) = -\nabla_{\qvec}\cdot (f \tilde{\Vvec}).
\end{equation}


\begin{remark}
    In common sense, $f\tilde{\Vvec}$ can be regarded as flux of $\qvec$ in the configurational space. 
\end{remark}

Moreover,  in the absence of external forces, another assumption is that the center of mass of the polymer chain $\rvec_c = \frac{1}2 (\rvec_1 + \rvec_2)$ moves along with the solution velocity at each $\x$, namely, the microvelocity of mass center equals to the macroscopic induced velocity \cite{bird1987}. 
According to the Cauchy-Born relation, it yields that 
\begin{equation}\label{eq_rc}
\dot{\rvec_c} = (\nabla_{\x} \uvec)^T \cdot \rvec_c.
\end{equation}
For mathematical simplicity, we set $\rvec_c = \bm 0$ in the Lagrangian coordinate $\Qvec$ of the configurational space at $t=0$, due to Eq. \eqref{eq_rc}, we obtain that $\rvec_c \equiv \bm 0$ for all time $t$.




\subsection{Model derivation}

In the following, we will derive the unknown terms $\bm\tau_s$, $\bm\tau_p$ and $ \tilde{\Vvec}$ by energy variational method. 
The total energetic functional  is defined as the sum of the kinetic energy of the fluid on the macroscopic scale,  the mixing energy and the spring potential energy of the polymers on the microscopic scale:
\begin{equation}
\begin{aligned}
& E_{total} =  E_{mac} +E_{mic}\\
& \quad = \int_{\Omega} \frac{1}{2} \rho |\uvec|^2 + \lambda_p\bigg[ \int_{\mathbb{R}^d} (k_B T f (\ln (f/f_{\infty}) -1)+\Psi f) \dd \qvec \bigg] \ \dd \x,  
\end{aligned}
\end{equation}
where $\Psi = \Psi(\qvec)$ is the spring potential energy functional,  $\lambda_p$ is a parameter related to the polymer density, $f_{\infty}$ is the distribution in the reference state, which is a constant.  Notice that due to the assumptions given in Section \ref{assulaw},  $E_{total}$ depends on the relative position $\qvec$ of two beads instead of $\rvec_1,\rvec_2$.
%
Then the  chemical potential is defined according to the energetic functional as follows:
$$
\mu_f = \frac{\delta E_{total}}{\delta f} = \lambda_p [k_B T \ln (f/f_\infty) +\Psi].
$$

The dissipative functional consists of fluid friction on the macroscale, and dissipation induced by the relative friction of microscopic particles to the macroscopic
flow 
 
\begin{equation}\label{dissipationdef}
\begin{aligned}
    \Delta &= \int_{\Omega} 2\eta_s |\bm D_{\eta}|^2 d\x  +    \lambda_p \zeta \int_{\Omega} \int_{\mathbb{R}^d}   \int_{\mathbb{R}^d}  \sum_{i=1}^2 f  |\dot{\rvec}_i - (\nabla_{\x} \uvec)^T \cdot \rvec_i |^2  \dd \rvec_1 \dd \rvec_2 \dd \x\\
\end{aligned}
\end{equation}
where $\eta_s$ is the viscosity of the fluid, $\bm D_{\eta} = \frac{1}{2} (\nabla_{\x} \uvec + (\nabla_{\x} \uvec)^T)$ is the strain rate, $\lambda_p$ and $\zeta$ are constants 
related to the polymer density and polymer relaxation time, respectively.
Since the Jacobian relation for the configurational vectors reads as follows \cite{bird1987}, 
$$
\bigg|  \frac{\partial(\rvec_1, \rvec_2)}{\partial (\rvec_c, \qvec)}  \bigg| = 1
$$
and $\rvec_c \equiv 0$ for all time $t$, by simple calculation, the dissipative functional can be rewritten as 
\begin{equation}\label{dissipationdef2}
\begin{aligned}
    \Delta &= \int_{\Omega} 2\eta_s |\bm D_{\eta}|^2 d\x   +  \frac{  \lambda_p \zeta } 2 \int_{\Omega} \int_{\mathbb{R}^d}   \int_{\mathbb{R}^d}  f  |\Vvec-(\nabla_{\x} \uvec)^T \cdot \qvec|^2  \delta(\rvec_c) \dd \rvec_c \dd \qvec \dd \x\\
     &= \int_{\Omega} 2\eta_s |\bm D_{\eta}|^2 d\x +  \frac{  \lambda_p \zeta } 2 \int_{\Omega} \int_{\mathbb{R}^d}    f  |\Vvec-(\nabla_{\x} \uvec)^T \cdot \qvec|^2   \dd \qvec \dd \x,\\
\end{aligned}
\end{equation}
where $ \delta(\rvec_c)$ is a Dirac function with respect to $\rvec_c$.

The energy dissipative law states that without an external force acting on the system, the changing rate of total energy equals to the dissipation \cite{xu2019three,eisenberg2010energy,xu2014energetic}:
\begin{equation}\label{energylaw}
    \frac{d}{dt} E_{total} = -\Delta.
\end{equation}
By the definition of total energy functional, the left-hand side of (\ref{energylaw}) is written as
\begin{equation}\label{dEdt}
    \frac{d}{dt} E_{total} = I_1+  I_2
\end{equation}
The first term $I_1$ is given by
\begin{equation}\label{I1}
\begin{aligned}
    I_1 &= \frac{d}{dt}E_{mac} = \int_{\Omega} \rho \uvec \cdot \uvec_t \mathrm{d}\x \\
    & = \int_{\Omega} \rho \uvec \cdot \bigg( \frac{D}{Dt} \uvec -\uvec\cdot \nabla_{\x} \uvec \bigg) \mathrm{d}\x\\
    & = \int_{\Omega} \uvec\cdot(-\rho \uvec \cdot \nabla_{\x} \uvec + \nabla_{\x} \cdot \bm \tau_s + \nabla_{\x} \cdot \bm \tau_p) \mathrm{d}\x\\
    & = \int_{\Omega} \nabla_{\x} \cdot (\rho \uvec) \frac{|\uvec|^2}{2} \mathrm{d}\x - \int_{\Omega} \nabla_{\x} \uvec :(\bm \tau_s + \bm \tau_p) \mathrm{d}\x\\
    & = \int_{\Omega} p(\nabla_{\x} \cdot \uvec) d\x - \int_{\Omega} \nabla_{\x} \uvec :(\bm \tau_s + \bm \tau_p) \mathrm{d}\x,
\end{aligned}
\end{equation}
where the pressure $p$ is introduced as a Lagrange multiplier for incompressibility.
%
The second term $I_2$ is given by
\begin{equation*}
    \begin{aligned}
    I_2
    & = \frac{d}{dt} E_{mic} =\int_{\Omega} \int_{\mathbb{R}^d} \mu_f f_t \dd \qvec \dd \x \\
     & = \int_{\Omega} \int_{\mathbb{R}^d} \mu_f \bigg(-\nabla_{\x} \cdot (\uvec f)  -  \nabla_{\qvec}\cdot( (\nabla_{\x} \uvec)^T \cdot \qvec f)
     -\nabla_{\qvec}\cdot(f \tilde{\Vvec})\bigg) \dd \qvec \dd \x \\ 
    & = \int_{\Omega} \int_{\mathbb{R}^d} \bigg(f\nabla_{\x}\mu_f\cdot\uvec +f\nabla_{\qvec}\mu_f\cdot((\nabla_{\x} \uvec)^T \cdot \qvec) +f \nabla_{\qvec}\mu_f \cdot  \tilde{\Vvec} \bigg)\dd \qvec \dd \x \\
    &= \lambda_p\int_{\Omega} \int_{\mathbb{R}^d}  k_B T\nabla_{\x} f\cdot\uvec \dd \qvec \dd \x+\int_{\Omega} \int_{\mathbb{R}^d} \bigg(\lambda_p(k_B T\nabla_{\qvec} f+f\nabla_{\qvec}\Psi)\cdot((\nabla_{\x} \uvec)^T \cdot \qvec)  +f \nabla_{\qvec}\mu_f\cdot \tilde{\Vvec} \bigg) \dd \qvec \dd \x\\
    &= \int_{\Omega} \int_{\mathbb{R}^d} \bigg(\underbrace{\lambda_p(k_B T\nabla_{\qvec} f+f\nabla_{\qvec}\Psi)\cdot( (\nabla_{\x} )\uvec)^T  \cdot \qvec}_{H_1} +\underbrace{f \nabla_{\qvec}\mu_f\cdot \tilde{\Vvec}}_{H_2}\bigg)\dd\qvec \dd\x, 
    \end{aligned}
\end{equation*}
where the equation \eqref{conservationf} and the incompressibility condition $\nabla_{\x} \cdot \uvec =0$ are used.

The first part $H_1$ in the integrand of $I_2$ is given by
%
\begin{eqnarray}
H_1 &=& \int_{\Omega} \lambda_p\int_{\mathbb{R}^d}  (k_B T  \nabla_{\qvec} f + f \nabla_{\qvec}\Psi) \cdot ((\nabla_{\x} \uvec)^T \cdot \qvec) \dd\qvec \dd \x\nonumber \\
&=& \lambda_p\int_{\Omega} \int_{\mathbb{R}^d}  -k_B T f \nabla_{\qvec} \cdot ((\nabla_{\x} \uvec)^T \cdot \qvec)
\dd \qvec \dd \x\nonumber + \int_{\Omega} \int_{\mathbb{R}^d}  \nabla_{\x} \uvec :[\nabla_{\qvec} \Psi \otimes \qvec ]f  \dd \qvec \dd\x\nonumber\\
& =& \lambda_p \int_{\Omega} \int_{\mathbb{R}^d}  \nabla_{\x} \uvec :[\nabla_{\qvec} \Psi \otimes \qvec]f  \dd \qvec \dd\x.\nonumber
\end{eqnarray}
Here, $\otimes$ denotes a tensor product and $\uvec\otimes\mathbf{v}$ is a matrix $(u_i v_j)$ for two vectors $\uvec$ and $\mathbf{v}$.
And here we use the fact that 
$$
\nabla_{\qvec} \cdot ( (\nabla_{\x} \uvec)^T \cdot \qvec) = \nabla_{\x} \cdot \uvec = 0.
$$

Thus, we have 
%
\begin{equation}\label{I2} 
I_2  = \int_{\Omega} \int_{\mathbb{R}^d} \lambda_p \nabla_{\x} \uvec :[\nabla_{\qvec} \Psi \otimes \qvec]f + f \nabla_{\qvec}\mu_f\cdot \tilde{\Vvec}
\dd \qvec \dd\x.
\end{equation}

Substituting Eqs. \eqref{I1} and \eqref{I2} into Eq. \eqref{dEdt}, we have
 
\begin{equation}
\begin{aligned}
   \frac{d}{dt} E_{total} &= I_1+   I_2\\
    & = \int_{\Omega} p(\nabla_{\x} \cdot \uvec) \dd\x \\
    & + \int_{\Omega} \nabla_{\x} \uvec :\bigg[\lambda_p \int_{\mathbb{R}^d}  [\nabla_{\qvec} \Psi \otimes \qvec ]f  \dd \qvec - 
    \bm \tau_s - \bm \tau_p\bigg] \dd\x+ \int_{\Omega} \int_{\mathbb{R}^d}   f \nabla_{\qvec}\mu_f\cdot \tilde{\Vvec} \dd\qvec \dd\x.
\end{aligned}
\end{equation}

Comparing with the energy dissipation law \eqref{energylaw} and the dissipation functional \eqref{dissipationdef2}, we obtain
\begin{equation}\label{jdef}
    \begin{aligned}
    &\bm \tau_s = 2\eta_s \bm D_{\eta}-p\bm I,\\
    &\bm \tau_p = \lambda_p \int_{\mathbb{R}^d}  [\nabla_{\qvec} \Psi \otimes \qvec ]f  \mathrm{d}\qvec,\\
     &\tilde{\Vvec} = -\frac{2}{\zeta \lambda_p} \nabla_{\qvec} \mu_f = -\frac{2}{\zeta}\bigg(k_B T \frac{1}{f} \nabla_{\qvec}  f + \nabla_{\qvec}\Psi\bigg).
    \end{aligned}
\end{equation}

In summary, the micro-macro model for polymeric fluids  with $N=2$ is
\begin{equation}\label{model}
\left\{
\begin{aligned}
&\rho(\uvec_t+\uvec \cdot \nabla_{\x} \uvec)+\nabla_{\x} p= \eta_s \Delta_{\x} \uvec+\nabla_{\x} \cdot \bm \tau_p,\\
&\bm \tau_p=\lambda_p\int_{\mathbb{R}^d}  [\nabla_{\qvec} \Psi \otimes \qvec]f  \mathrm{d}\qvec,\\
& \nabla_{\x}\cdot\uvec=0,\\
& f_t+ \uvec \cdot \nabla_{\x} f+ \nabla_{\qvec} \cdot((\nabla_{\x} \uvec)^T \cdot \qvec f) =  \frac{2}{\zeta}\nabla_{\qvec}\cdot(f\nabla_{\qvec} \Psi )+ \frac{2k_B T}{\zeta} \Delta_{\qvec} f,
\end{aligned}
\right.
\end{equation}

\begin{remark}
In this article, we only focus on the model with two beads. Actually, the model derivation via energy variational method can be directly applied to the multi-bead case (namely, Eq. \eqref{CPsystemwithf_1} with $N \geq 3$). 
%
%
%
%
The multi-bead model will be discussed in our upcoming works. 
\end{remark}

\subsection{Microscopic potential energy}

In this article, we compare the behaviors of polymer chains under conditions of  three different microscopic potential energies: the classical Hooekan potential, modified Morse potential and Elastic-plastic potential.


The classical Hooekan potential reads as follows
$$
\Psi(\qvec)=\frac{1}2H|\mathbf{q}|^2, 
$$
and the derivative of the potential $\Psi$, namely the so-called entropic force \cite{bird1987}, is given by
$$
\quad \nabla_{\mathbf{q}} \Psi(\qvec)=H \mathbf{q},
$$
where $H>0$ is the elastic constant. 

The Elastic-plastic potential assumes that when the end-to-end distance
between beads gets larger than a critical value,  
the interactions between two beads remain a constant value. 
This kind of abrupt change describes the behavior of bond breaking.  
A natural idea to construct this potential energy is to truncate the classical linear elastic spring potential.
And thus, Elastic-plastic potential is given by 
$$
\Psi(\qvec)=\left\{\begin{aligned}
&\frac{1}2H|\mathbf{q}|^2, \qquad \mbox{when} \quad |\mathbf{q}|\leq Q_0, \\
& \frac{1}2H Q_0^2,  \qquad \mbox{when} \quad |\mathbf{q}|>Q_0. 
\end{aligned}
\right.
$$
The corresponding entropic force is 
$$
\quad \nabla_{\mathbf{q}} \Psi(\mathbf{q})=\left\{\begin{aligned}
&H\mathbf{q}, \qquad \mbox{when} \quad |\mathbf{q}|\leq Q_0, \\
&0, \qquad \mbox{when} \quad |\mathbf{q}|>Q_0.
\end{aligned}
\right.
$$
Here we assume that when end-to-end distance $|\qvec|$ is small ($|\qvec|\leq Q_0$), it is the classical Hookean model. But when $|\qvec|$ gets larger to the truncation constant $Q_0$, the bond breaking occurs and the broken part moves along with the flow. And thus, when $|\qvec|> Q_0$, the potential equals to a constant.

\begin{figure}[!ht]
\centering
 \begin{overpic}[width=0.5\linewidth]{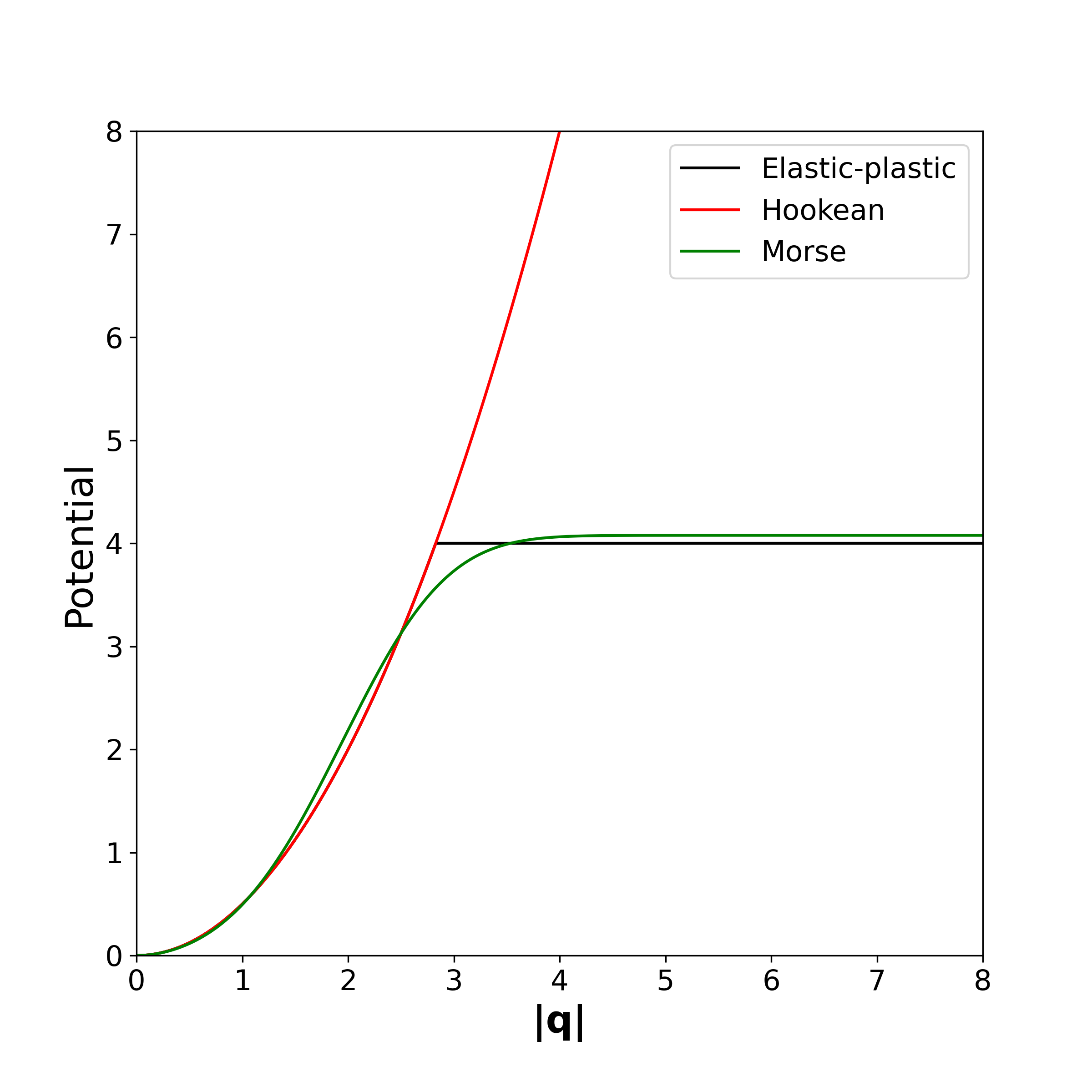}
 \end{overpic}
\caption{
Different potentials with respect to end-to-end vector distance $|\qvec|$. }  
\label{micromacro}
\end{figure}

Different from the classical elastic dumbbell model, if the effects of bond breaking are included, Morse potential is considered, which is given by:
%
$$
\Psi(r)=D\bigg(1-e^{-\beta(r-r_e)}\bigg)^2,
$$
where $r$ is the distance between two beads, $r_e$ is the equilibrium bond distance, $D$ is the well depth and $\beta$ controls the `width' of the potential. 
Here, we set the equilibrium bond distance to be zero, i.e., only the adhesion is considered. 
Based on the classical Morse potential, we use the modified Morse potential in this article such that it could fit the classical Hookean potential when $|\mathbf{q}|$ is small and the force decrease to zero around the breaking threshold of the Elastic-plastic potential  (see Figure \ref{micromacro}, the Elastic-plastic potential behaves like an abrupt version of the Morse potential).
The modified Morse potential reads as follows, 
$$
\Psi(\mathbf{q})=D\bigg(1-e^{-\beta|\mathbf{q}|-\alpha |\mathbf{q}|^3 }\bigg)^2,
$$
and the corresponding entropic force is given by
$$
\nabla_{\mathbf{q}}\Psi = 2D(1-e^{-\beta|\mathbf{q}|-\alpha|\mathbf{q}|^3})e^{-\beta|\mathbf{q}|-\alpha|\mathbf{q}|^3}\bigg(\beta \frac{\qvec}{|\qvec|} + 3\alpha \qvec |\qvec|\bigg).
$$
For simplicity, we still call this potential as Morse potential in this article.

%
%
%

In the rest of this article, we denote the micro-macro model \eqref{model} with microscopic Hookean, modified Morse, and Elastic-plastic potentials as Hookean, Morse, and Elastic-plastic models, respectively.

%


%
%

\subsection{Non-dimensionalization}

In this section, we present the dimensionless version of the micro-macro system \eqref{model}. 
To this end, we start from the dimensionless form of the energy dissipation law and introduce the dimensionless variables \cite{le2012micro}
 
\begin{equation}\label{dimensionlessvariable}
\tilde{\x} = \frac{\x}{\tilde{L}}, \quad \tilde{\qvec}=\frac{\qvec}{\tilde{L}}, \quad \tilde{\uvec} = \frac{\uvec}{\tilde{U}}, \quad \tilde{t} = \frac{t}{\tilde{T}}, \quad \tilde{T} = \frac{\tilde{L}}{\tilde{U}}. 
\end{equation}
Here $\tilde{L}= \sqrt{k_B T/H}$, $\tilde{T}$, $\tilde{U}$ are the characteristic length, time and velocity. We define
$\lambda = \zeta/4H$, $\eta_p = \lambda_p k_B T \lambda$ and $\eta = \eta_s + \eta_p$, where $\eta_p$ is the parameter related to the polymer viscosity and $\eta$ the total fluid viscosity.
 
For convenience, the tilde symbol has been removed in dimensionless quantities here. 
Thus, we obtain the dimensionless version of the energy dissipation law  
\begin{equation}\label{dissipationlaw_nondim}
\begin{aligned}
&\frac{d}{dt} \int_{\Omega} \frac{1}{2} \mathrm{Re} |\uvec|^2 + \frac{\epsilon_p}{\mathrm{Wi}} \bigg[ \int_{\mathbb{R}^d} ( f (\ln f - C ) +\Psi f) \mathrm{d}\qvec \bigg] \ \mathrm{d}\x\\
& \quad =-\int_{\Omega} \frac{1}{2} \tilde{\eta}_s |\nabla \uvec|^2 \dd\x - 2 \epsilon_p \int_{\Omega} \int_{\mathbb{R}^d} f   \bigg|\Vvec - (\nabla_{\x} \uvec)^T \cdot \qvec\bigg|^2 \dd\qvec \dd\x.
\end{aligned}
\end{equation}
with the dimensionless parameters
$$
\mathrm{Re} = \frac{\rho \tilde{U} \tilde{L}}{\eta}, \quad \mathrm{Wi} = \frac{\lambda \tilde{U}  }{\tilde{L}}, \quad \tilde{\eta}_s = \frac{\eta_s}{\eta}, \quad \epsilon_p = \frac{\eta_p}{\eta}.
$$


The dimensionless system of \eqref{model} is given by 
\begin{equation}\label{nondimensionmodel}
\left\{
\begin{aligned}
& \mathrm{Re} (\uvec_t+\uvec \cdot \nabla_{\x} \uvec)+\nabla_{\x} p= \tilde{\eta}_s \Delta_{\x} \uvec+\nabla_{\x} \cdot \bm \tau_p,\\
&\bm \tau_p= \frac{\varepsilon_p}{\mathrm{Wi}} \int_{\mathbb{R}^d}  f \nabla_{\qvec} \Psi \otimes \qvec  \mathrm{d}\qvec,\\
& \nabla_{\x}\cdot\uvec=0,\\
& f_t+ \uvec \cdot \nabla_{\x} f+ \nabla_{\qvec} \cdot((\nabla_{\x} \uvec)^T \cdot \qvec f) =  \frac{1}{2 \mathrm{Wi}}[\nabla_{\qvec}\cdot(f\nabla_{\qvec} \Psi )+  \Delta_{\qvec} f ].
\end{aligned}
\right.
\end{equation}
After the non-dimensionalization, the Hookean potential becomes
$$
\Psi(\qvec) = \frac{1}{2}  |\qvec|^2, 
$$
 Morse potential becomes
$$
\Psi(\mathbf{q})=\frac{1}{2\beta^2} \bigg(1-e^{-\beta|\mathbf{q}|-\alpha |\mathbf{q}|^3 }\bigg)^2.
$$
and Elastic-plastic potential becomes
$$
\Psi(\qvec)=\left\{\begin{aligned}
&\frac{1}2|\mathbf{q}|^2, \quad |\mathbf{q}|\leq Q_0, \\
& \frac{1}2 Q_0^2,  \quad |\mathbf{q}|>Q_0.
\end{aligned}
\right.
$$

\section{Numerical Method}

In this article, we utilize the scheme proposed in Ref. \cite{bao2022} for numerical simulations, by combining a finite element discretization for the macroscopic fluid equation \cite{bao2021, becker2008,chenrui2015} with a deterministic particle method for the microscopic Fokker-Planck equation \cite{wangEVI}.



\subsection{Deterministic particle method}

Due to Eq. \eqref{int_n} and definition of $f$, for fixed $\x$, $f(\x, \qvec, t)$ is approximated by a particle approximation: 
\begin{equation}\label{approxf}
f(\x, \qvec, t) \approx f_N (\x, \qvec, t) = \sum_{i=1}^{N_p}  \omega_i(\x, t)  \delta(\qvec - \bar{\qvec}_i(\x, t)).
\end{equation}
Here, $\{\bar{\qvec}^{\x}_i (t) \}_{i=1}^{N_p}$ is a set of particles and can be viewed as representative particles, representing information of the number density distribution $f(\x, \qvec, t)$ at position $\x$ and time $t$. 
$N_p$ is the number of particles. $\omega_i(\x, t)$ is the weight of the corresponding particle satisfying $\sum_{i=1}^{N_p}\omega_i(\x, t) = 1$. 
As in the former work \cite{bao2022}, we also fix $\omega_i(\x, t) = \frac{1}{N_p}$ in this article, namely, all particles are equally weighted.

\begin{remark}\label{rmk：qbar}
Since the particle method is a coarse-grained method, each particle $\bar{\qvec}_i$ is not each real microscopic polymer chain. 
However, the distribution information of particles $\{\bar{\qvec}_i(\x, t)\}_{i=1}^{N_p}$ reveals the information of a set of corresponding real polymer chains. And thus, we can investigate the lengths and directions of polymer chains in the fluid by investigating the lengths and directions of the coarse grained particles. 
\end{remark}



The micro-macro system with particles $\{\bar{\qvec}_i(\x, t) \}_{i=1}^{N_p}$ can be derived by utilizing the discrete energetic variational approach \cite{wang2020jcp, wangEVI, bao2022}. The discrete energetic variational approach is analogous to energetic variational approaches in a semidiscrete level. 
Firstly, substitute the approximation \eqref{approxf} into the continuous energy-dissipation law \eqref{dissipationlaw_nondim}, a discrete energy-dissipation law in terms of $\{\bar{\qvec}_i \}_{i=1}^{N_p}$ and the macroscopic flow can be obtained:
\begin{equation}\label{discreteenergylaw}
\begin{aligned}
\frac {d}{dt}\mathcal{F}_h \left[ \{\bar{\qvec}_i\}_{i=1}^{N_p}, \x \right] = -\mathcal{D}_h \left[ \{\bar{\qvec}_i\}_{i=1}^{N_p}, \{\dot{\bar{\qvec}}_i\}_{i=1}^{N_p}, \x, \uvec  \right],
\end{aligned}
\end{equation}
where 
\begin{equation}
\label{discreteenergylaw2}
\begin{aligned}
\mathcal{F}_h \left[ \{\bar{\qvec}_i\}_{i=1}^{N_p}, \x \right] = \int_{\Omega} \frac{1}{2} \mathrm{Re} |\uvec|^2 +\frac{\epsilon_p}{ \mathrm{Wi} } \frac{1}{N_p} \sum_{i=1}^{N_p} \bigg{[} \ln \left(\frac{1}{N_p} \sum_{j=1}^{N_p} K_h(\bar{\qvec}_i - \bar{\qvec}_j )\right) +\Psi(\bar{\qvec}_i)\bigg{]} \dd \x.\\
\end{aligned}
\end{equation}
and
\begin{equation}
\label{discreteenergylaw3}
\begin{aligned}
&\mathcal{D}_h \left[ \{\bar{\qvec}_i\}_{i=1}^{N_p}, \{\dot{\bar{\qvec}}_i\}_{i=1}^{N_p}, \x, \uvec  \right]\\
&=\int_{\Omega} \frac 1 2\tilde{\eta}_s |\nabla_{\x} \uvec|^2 + 2\epsilon_p \frac{1}{N_p} \sum_{i=1}^{N_p} |\bar{\Vvec}_i|^2 \dd \x \\
&=\int_{\Omega} \frac 1 2\tilde{\eta}_s |\nabla_{\x} \uvec|^2 + 2\epsilon_p \frac{1}{N_p} \sum_{i=1}^{N_p} |\dot{\bar{\qvec}}_i - (\nabla_{\x} \uvec)^T \cdot \bar{\qvec}_i|^2 \dd \x. 
\end{aligned}
\end{equation}
For the discrete energy,  the term $\ln(f_N)$ is  replaced  by $\ln (K_h\ast f_N)$ with a kernel function  $K_h$ for a kernel regularization \cite{Carrilloblob, bao2022} such that
\begin{eqnarray}
K_h\ast f_N(:, \qvec, :)
&=&\int K_h(\qvec-\bm{p}) f_N(:, \bm{p}, :) \dd {\bm p} =\frac{1}{N_p} \sum_{j=1}^{N_p} K_h(\qvec-\bar{\qvec}_j).
\end{eqnarray}
A typical choice of $K_{h}$ is the Gaussian kernel, given by
$$
K_{h}(\qvec_1, \qvec_2)=\frac{1}{(\sqrt{2\pi}h_p)^d}\exp\left(-\frac{|\qvec_1-\qvec_2|^2}{2h_p^2}\right).
$$
Here, $h_p$ is the kernel bandwidth which controls the inter-particle
distances and $d$ is the dimension of the space. In this article, we choose the parameters same as those in Ref. \cite{bao2022}.

Then the system of particles $\{\bar{\qvec}_i \}_{i=1}^{N_p}$ can be derived by performing the EnVarA in terms of $\bar{\qvec}_i$ and $\dot{\bar{\qvec}}_i$:
$$
\frac{\delta (\frac{1}2 \mathcal{D}_h)}{\delta \dot{\bar{\qvec}}_i} = -\frac{\delta \mathcal{F}_h}{\delta \bar{\qvec}_i}. 
$$
%
%
The final micro-macro system with particle approximation can be summarized as
\begin{equation}\label{finalmodel_dimensionless}
\left\{
\begin{aligned}
&\mathrm{Re}(\uvec_t+\uvec \cdot \nabla_{\x} \uvec) + \nabla_{\x} p= \tilde{\eta}_s \Delta_{\x} \uvec+\nabla_{\x} \cdot {\bm \tau_p},\\
&\bm \tau_p(\x, t) =\frac{\epsilon_p}{\mathrm{Wi}} \frac{1}{N_p} \sum_{i=1}^{N_p} \nabla_{\bar{\qvec}_i} \Psi(\bar{\qvec}_i (\x, t))\otimes \bar{\qvec}_i(\x, t),\\
& \nabla_{\x} \cdot\uvec=0, \\
& \pp_t \bar{\qvec}_i + (\uvec \cdot \nabla_{\x}) \bar{\qvec}_i - (\nabla_{\x} \uvec)^T\cdot \bar{\qvec}_i \\
  &\qquad =- \frac{1}{2\mathrm{Wi}} \bigg[\bigg(\frac{\sum_{j=1}^{N_p}  \nabla_{\bar{\qvec}_i} K_h(\bar{\qvec}_i- \bar{\qvec}_j)}{\sum_{j = 1}^{N_p} K_h(\bar{\qvec}_i-\bar{\qvec}_j) } + \sum_{k=1}^{N_p} \frac{\nabla_{\bar{\qvec}_i} K_h(\bar{\qvec}_k-\bar{\qvec}_i )}{\sum_{j=1}^{N_p} K_h(\bar{\qvec}_k-\bar{\qvec}_j)}\bigg) + \nabla_{\bar{\qvec}_i} \Psi(\bar{\qvec}_i) \bigg].
\end{aligned}
\right.
\end{equation}

\subsection{Numerical scheme}

 To solve the system \eqref{finalmodel_dimensionless} numerically, some decoupled schemes are applied at each step. Precisely, we apply the following algorithm proposed in Ref.  \cite{bao2022} for temporal discretization. 

Given the time step size $\Delta t$, the initial conditions $\{\bar{\qvec}_{\x_{\alpha}, i}^{0}\}_{i=1}^N$, $\uvec^0$, and $p^0$ at each node $\x_{\alpha}$ ($\alpha = 1, 2, \ldots N_x$), having computed for $\{\bar{\qvec}_i^{n}\}_{i=1}^{N_p}$, $\uvec^n$, $\bm \tau_p^n$ and $p^n$ for $n>0$, we compute  $\{\bar{\qvec}_{\x_{\alpha}, i}^{n+1}\}_{i=1}^{N_p}$, $\uvec^{n+1}$, and $p^{n+1}$ by the following algorithm:

\noindent{\bf Step 1:} Treating the stress tensor explicitly, solve the macroscopic flow equation via
\begin{equation}\label{Eq_u_n}
\begin{aligned}
  & \mathrm{Re} \bigg(\frac{\uvec^{n+1} - \uvec^n}{\Delta t} + \uvec^n \cdot \nabla_{\x} \uvec^{n+1}\bigg) + \nabla_{\x} p^{n+1}  = \tilde{\eta}_s \Delta_{\x} \uvec^{n+1} + \nabla_{\x} \cdot \bm\tau_p^n, \\
& \nabla_{\x} \cdot \uvec^{n+1} = 0. \\
\end{aligned}
\end{equation}

The equation (\ref{Eq_u_n}) can be solved by a standard velocity-correction projection method \cite{Chorin1969, Guermond2006}: 
    \begin{itemize}
      \item Step 1.1: Find $\tilde{\uvec}^{n+1}$, such that 
  \begin{equation*}
   \mathrm{Re} \bigg( \frac{\tilde{\uvec}^{n+1}-\uvec^{n}}{\Delta t}  +\uvec^n \cdot \nabla_{\x} \tilde{\uvec}^{n+1}  \bigg) + \nabla_{\x} p^n = \tilde{\eta}_s \Delta_{\x} \tilde{\uvec}^{n+1} + \nabla_{\x} \cdot \bm \tau_p^n.
  \end{equation*} \\
      \item Step 1.2: Find $p^{n+1}$, such that, 
  \begin{equation*}
  \begin{aligned}
  &\mathrm{Re} \frac{\uvec^{n+1}-\tilde{\uvec}^{n+1}}{\Delta t} + \nabla_{\x} (p^{n+1} - p^n)=0,\\
  &\nabla_{\x} \cdot \uvec^{n+1} = 0.
  \end{aligned}
  \end{equation*}
  Then we obtain 
  $$ \uvec_h^{n+1} = \tilde{\uvec}_h^{n+1} - \Delta t \nabla_{\x} (p_h^{n+1}-p_h^n).$$
\end{itemize}

\noindent{\bf Step 2}: Within the values $\uvec_h^{n+1}$, we solve the microscopic equation in \eqref{finalmodel_dimensionless} by an update-and-project approach. 
Precisely, we use a Lagrangian approach to compute the convection term \cite{Halin1998} and use an operator splitting approach.
Initially, we assign ensemble of particles $\{\bar{\qvec}_{\x_\alpha,  i}^n \}_{i=1}^{N_p}$ to each node $\x_{\alpha}$ ($\alpha = 1, 2, \ldots N_x$). 
Within the values $\uvec_h^{n+1}$, we solve the microscopic equation by the following two steps:
  \begin{itemize}
      \item  Step 2.1: At each node $\x_{\alpha}$, solve the microscopic equation in \eqref{finalmodel_dimensionless} without the convection term $\uvec^{n+1} \cdot \nabla_{\x} \qvec_i$ by
    \begin{equation}\label{ODE_q_1}
    \begin{aligned}
         \frac{\bar{\qvec}_i^{n+1, *}-\bar{\qvec}_i^{n}}{\Delta t} & =- \frac{1}{2\mathrm{Wi}} \bigg[\bigg(\frac{\sum_{j=1}^{N_p}  \nabla_{\bar{\qvec}_i} K_h(\bar{\qvec}_i^{n+1, *}- \bar{\qvec}_j^{n+1, *})}{\sum_{j = 1}^{N_p} K_h(\bar{\qvec}_i^{n+1, *}-\bar{\qvec}_j^{n+1, *}) } \\
  & \qquad  \quad + \sum_{k=1}^{N_p} \frac{\nabla_{\bar{\qvec}_i} K_h(\bar{\qvec}_k^{n+1, *}-\bar{\qvec}_i ^{n+1, *})}{\sum_{j=1}^{N_p} K_h(\bar{\qvec}_k^{n+1, *}-\bar{\qvec}_j^{n+1, *})}\bigg) + \nabla_{\bar{\qvec}_i} \Psi(\bar{\qvec}_i^{n+1, *}) \bigg]. \\     
        & \bar{\qvec}_i^{n+1, **} = (I + \Delta t (\nabla_{\x} \uvec^{n+1}) )^T \cdot \bar{\qvec}_i^{n+1, *}.  
    \end{aligned}
    \end{equation}\\
    \item Step 2.2: To deal with the convection term, we view each node $\x_{\alpha}$ as a Lagrangian particle, and update it according to the Eulerian velocity field $\uvec_h^{n+1}$ at each node
\begin{equation}
\tilde{\x}_{\alpha} = \x_{\alpha} + \Delta t  (\uvec^{n+1}_h |_{\x_{\alpha}}), \quad \alpha = 1, 2, \ldots N_x.
\end{equation}
Hence, $\{ \bar{\qvec}_{\tilde{\x}_{\alpha}, i}^{n+1, **} \}$ is an ensemble of samples at the new point $\tilde{\x}_{\alpha}$. In order to obtain $\bar{\qvec}_{\x_{\alpha}, i}^{n+1}$ at $\x_{\alpha}$, for each $i$,     a
linear interpolation is adopted to get $\bar{\qvec}_{\x_{\alpha}, i}^{n+1}$ from $\bar{\qvec}_{\tilde{\x}_{\alpha}, i}^{n+1, **}$.  \\
\end{itemize}

And thus, the updated values of the stress $\bm \tau_p^{n+1}$ at each node, denoted as $\{\bm \tau_{p,\x_{\alpha}}^{n+1}\}_{\alpha = 1}^{N_x}$, can be directly computed through the second equation of Eq. \eqref{finalmodel_dimensionless}.

\begin{remark}
This numerical method was first proposed in Ref. \cite{bao2022} and has been studied for Hookean and FENE models. 
The accuracy of this numerical method has been evaluated with a small number of particles. 
Moreover, the impact of the number of particles has also been investigated, revealing that a good numerical result can be achieved with particle number $N_p = 200$. Thus, we set particle number $N_p=200$ for all numerical experiments in this article.
    
\end{remark}

\begin{remark}
Actually, to simulate the micro-macro system \eqref{finalmodel_dimensionless}, only particles $\{\bar{\qvec}_i\}_{i=1}^{N_p}$ needs to be computed at each time step. 
And it has been numerically validated \cite{bao2022} that effective numerical results can be achieved with particle number $N_p = 200$. 
Compared with the stochastic methods, such as the CONNFFESSIT method \cite{Laso1993}, Lagrangian particle method \cite{Halin1998} and Brownian configuration fields method \cite{Hulsen1997}, the computational cost can be largely reduced and numerical results show few stochastic fluctuations, see Ref. \cite{bao2022} and references therein. 
\end{remark}

\section{Results and Discussions}

In this section, we investigate the irreversible bond breaking induced by shear flow conditions, via simulating the micro-macro model \eqref{finalmodel_dimensionless} with Morse and Elastic-plastic potentials. 
We also simulate the classical Hookean model as a comparison. 
Precisely, numerical results are devoted to investigating the effects of microscopic bond breaking on the stresses of the macroscopic fluid by investigating the configurations of polymer chains. 
In this article, the distribution of polymer chains is approximated by a deterministic particle method, and thus, one can obtain the configurations of real polymer chains by investigating the deterministic particles  $\{\bar{\qvec}_i(\x, t)\}_{i=1}^{N_p}$.  In the simulation,  the following parameters are adopted
\begin{equation*}
\tilde{\eta}_s=0.111,~ \epsilon_p=1-\tilde{\eta}_s=0.889,  ~ \Delta t =0.002, ~ Q_0 = \sqrt{8},~\beta=0.3502,~\alpha = 0.07734.
\end{equation*}


\subsection{Simple flow with constant shear rate }
\begin{figure}[htpb]
\centering
\includegraphics[width = 0.5\linewidth]{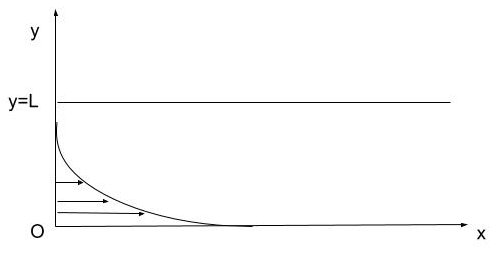}
\caption{Schematic representation of the initial simple shear flow.}
\label{initialflow}
\end{figure}

We first consider the simple shear flow, where the fluid is enclosed between two parallel planes of an infinite length separated by a distance $L$. At $t = 0$, the lower plane starts to move
in the positive $x$ direction with a constant velocity $U_{bottom}$, as shown in Figure \ref{initialflow}. 
Such a model is typically obtained by considering a flow in a rheometer, between two cylinders, and taking the limit of large radii for both the inner and outer cylinders. It is natural to assume that the flow is laminar, that is,
the velocity  $\uvec(\x, t) = (u(y, t), 0)$ is in the $x$-direction, and only depends on the $y$-variable with $\x = (x, y)$. Obviously, the velocity field automatically satisfies the incompressible condition $\nabla_{\x} \cdot \uvec = 0$. Moreover, we assume that $\qvec$ also only depends on $y$, so we obtain $\uvec\cdot \nabla_{\x} \qvec = 0$. 

Owing to these assumptions, the micro-macro system \eqref{finalmodel_dimensionless} can be simplified as
\begin{equation}\label{finalmodel1D}
\left\{
\begin{aligned}
& \mathrm{Re} \frac{\partial u}{\partial t} = \tilde{\eta}_s u_{yy} + \frac{\partial \tau_{p21}}{\partial y},\\
& \tau_{p21} = \frac{\epsilon_p}{\mathrm{Wi}} \frac{1}{N_p} \sum_{i=1}^{N_p} [\nabla_{\qvec} \Psi(\bar{\qvec}_i)]_{2} [\bar{\qvec}_{i}]_1,\\
& \frac{\partial [\bar{\qvec}_{i}]_1 }{\partial t}  - \frac{\partial u}{\partial y} [\bar{\qvec}_{i}]_2  = - \frac{1}{2\mathrm{Wi}} \bigg[\bigg(\frac{\sum_{j=1}^{N_p}  [\nabla_{\qvec_i} K_h(\bar{\qvec}_i- \bar{\qvec}_j)]_1}{\sum_{j = 1}^{N_p} K_h(\bar{\qvec}_i-\bar{\qvec}_j) } \\
& \qquad  \quad + \sum_{k=1}^{N_p} \frac{[\nabla_{\qvec_i} K_h(\bar{\qvec}_k-\bar{\qvec}_i )]_1}{\sum_{j=1}^{N_p} K_h(\bar{\qvec}_k-\bar{\qvec}_j)}\bigg) + [\nabla_{\qvec_i} \Psi(\bar{\qvec}_i)]_1 \bigg],\\
& \frac{\partial [\bar{\qvec}_{i}]_2 }{\partial t}
= - \frac{1}{2\mathrm{Wi}} \bigg[\bigg(\frac{\sum_{j=1}^{N_p}  [\nabla_{\qvec_i} K_h(\bar{\qvec}_i- \bar{\qvec}_j)]_2}{\sum_{j = 1}^{N_p} K_h(\bar{\qvec}_i-\bar{\qvec}_j) }+ \sum_{k=1}^{N_p} \frac{[\nabla_{\qvec_i} K_h(\bar{\qvec}_k-\bar{\qvec}_i )]_2}{\sum_{j=1}^{N_p} K_h(\bar{\qvec}_k-\bar{\qvec}_j)}\bigg)
 + [\nabla_{\qvec} \Psi(\bar{\qvec}_i)]_2 \bigg].
\end{aligned}
\right.
\end{equation}
%
%
Here, we assume that the dimension of the configurational space $d=2$ and the notation $[\bm g]_l$ means the $l$-th component of the vector $\bm g$. In the above dimensionless system, the domain is set as $[0,1]$ and the velocity of the lower plane is set as $U_{bottom} = 1$.

\subsubsection{Velocity  }
\begin{figure*}[htpb]
   \centering
\includegraphics[width = 0.9\linewidth]{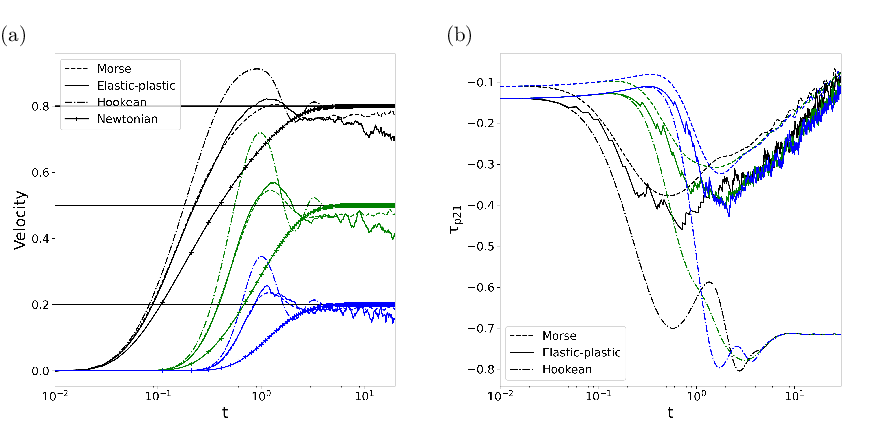}
   \caption{ The time evolution of (a) velocity, and (b) shear stress $\tau_{p21}$, with Wi, Re=1 for different models on three locations: $y=0.2$ (black), $y=0.5$ (green), $y=0.8$(blue). Time is in a logarithmic scale.} 
   \label{uandtaut_plot}
\end{figure*}

We begin by examining the time evolution of velocity for macroscopic fluids using a logarithmic scale, with the above three microscopic potential energies. Non-Newtonian fluids often exhibit unique behaviors, such as the rod-climbing effect, the open siphon effect \cite{owens2002}, and the overshoot phenomenon \cite{Laso1993}. As illustrated in Figure \ref{uandtaut_plot} (a), Newtonian fluids display a steadily increasing velocity profile, eventually reaching a stable state. In contrast, the Hookean, Morse, and Elastic-plastic models all exhibit velocities that exceed the steady-state value, displaying an overshoot phenomenon. This behavior suggests that the Morse and Elastic-plastic models share common characteristics with classical non-Newtonian fluids.

Next, we investigate the impact of bond breaking on the velocity evolution. When bond breaking occurs, the entropic force $\nabla_{\mathbf{q}}\Psi$ becomes negligible, causing the broken polymer chains to become unbounded and move along with the macroscopic fluid. Consequently, the velocity of the broken segments approaches that of a Newtonian fluid. As shown in Figure \ref{uandtaut_plot} (a), the velocity curves for the Morse and Elastic-plastic models fall between those of the classical Hookean model and the Newtonian fluid.

\vspace{0.5cm}

\subsubsection{Shear stress }

Now we explore the time evolution of stress $\bm \tau_p$ due to the microscopic polymer chains. Figure \ref{uandtaut_plot} (b) illustrates the shear stress $\tau_{p21}$ as a function of time $t$ for the three models. As observed in Figure \ref{uandtaut_plot} (b), the stress profiles exhibit more rapid changes at initial stage for locations closer to the moving plane (i.e., the black curves for y=0.2), but the profiles for all locations eventually converge to similar values, consistent with previous numerical findings \cite{Laso1993, XuSPH2014}. Notably, the shear stresses in the classical Hookean and Elastic-plastic models initially coincide which is expected since the two potentials are the same when the polymer norm $|\qvec|$ is small. Over time, bond breaking occurs, causing the shear stresses $\tau_{p21}$ in the Elastic-plastic and Morse models to approach zero. In contrast, the stress $\tau_{p21}$ for the  classical Hookean model tend to some finite value as time evolves. In the following, we investigate this distinction of $\tau_{p21}$ in two perspectives:  particle distribution and energy evolution.

First, we examine the influence of polymer distribution on the shear stress of macroscopic fluids. In this study, the distribution of particles $\{\bar{\qvec}_i\}_{i=1}^{N_p}$ approximates the distribution of polymer chains, see Remark \ref{rmk：qbar}. The particle property is mainly reflected by two quantities: particle elongation measured by the norm $|\bar{\qvec}_i|$, and particle orientation measured by $\sin\theta_i$, where $\theta_i$ is the direction of vector $\bar{\qvec}_i$. Thus, we investigate the distributions of particles in terms of $\sin\theta$ and norm $|\bar{\qvec}|$, using kernel density estimation (KDE) with a Gaussian kernel.

\begin{figure*}[htpb]
   \centering
     \includegraphics[width = 0.9\linewidth]{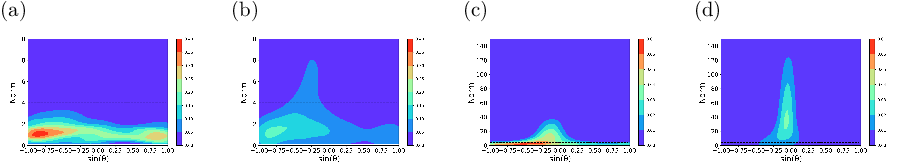}
   \caption{  Morse model: Contour plot of the particle distribution in terms of $\sin(\theta)$ and norm $|\bar{\qvec}|$ with Wi, Re=1 at different times (from (a) to (d): t=0.5, 5, 10, 30). }
   \label{morse_KDE_plot}
\end{figure*}

\begin{figure*}[htpb]
   \centering
      \includegraphics[width = 0.9\linewidth]{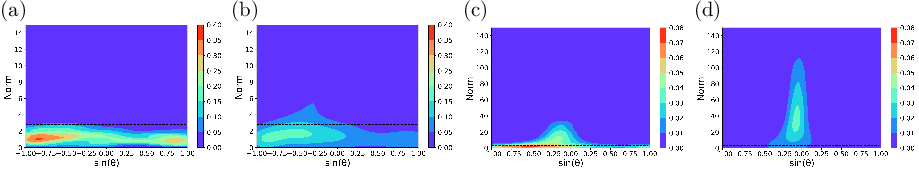}
   \caption{ Elastic-plastic model: Contour plot of particle distribution as a function of $\sin \theta$ and norm $|\bar{\qvec}|$ with Wi, Re=1 at different times (from (a) to (d): t=0.5, 5, 10, 30).} 
   \label{plastic_KDE_plot}
\end{figure*}

For the Morse and Elastic-plastic models, Figures \ref{morse_KDE_plot} and \ref{plastic_KDE_plot} present contour plots of particle distributions across the entire domain, in terms of $\sin\theta$ and norm $|\bar{\qvec}|$ with Wi = 1 and Re=1. The black dashed lines in these figures denote the truncations, $|\bar{\qvec}|=Q_0$ for the Elastic-plastic model and $|\bar{\qvec}|=4$ for the Morse model. In these contour plots, particles located above the black dashed line are approximately unbounded, corresponding to the bond breaking occurrence, see Figure \ref{micromacro}. Figures \ref{morse_KDE_plot} and \ref{plastic_KDE_plot} demonstrate that, for Morse and Elastic-plastic models, as the plane moves at a constant velocity, polymer chains elongate in the flow direction (resulting in longer $|\bar{\qvec}|$) and tend to align with the flow direction (indicated by $\sin \theta$ approaching zero). Specifically, at $t=5$, the contour plots of particle distribution are wider at the bottom and narrower at the top, forming a "pear" shape. This shape suggests that elongated particles (with larger $|\bar{\qvec}_i|$) tend to rotate in the flow direction (with $\sin \theta$ approaching zero). This phenomenon can be attributed to entropic forces: particles with lower entropic forces are more likely to rotate with the flow. For the Morse model, the magnitude of entropic force (see the slope of tangent lines on the potential in Figure \ref{micromacro}) decreases as the particle norm $|\bar{\qvec}_i|$ increases beyond a threshold (the inflection point). Similarly, for the Elastic-plastic model, when the particle norm exceeds $Q_0$, the entropic force becomes zero. Consequently, elongated particles tend to rotate with the flow, creating the "pear" shape observed in the contour plots. Furthermore, as time evolves, for Morse and Elastic-plastic models, the contour plots transit from a "pear" shape (at $t=5$) to a "stripe" shape (at $t=30$), which narrows uniformly at the top and bottom. This suggests that over time, nearly all particles elongate and rotate with the flow, with Wi = 1 and Re=1. Nearly the entire "stripe" shape in the contour plot ends up above the dashed line (at $t=30$), indicating that bond breaking occurs for nearly all particles and consequently the shear stress $\tau_{p21}$ is reduced. 

\begin{figure*}[htpb]
   \centering
     \includegraphics[width = 0.9\linewidth]{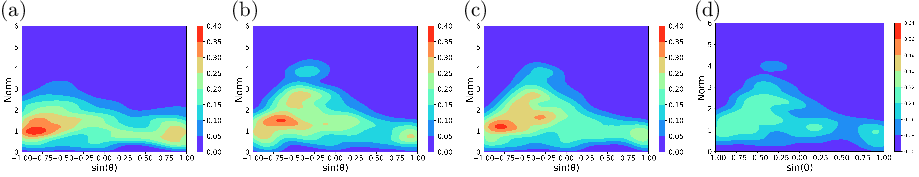}
        \caption{Hookean model: Contour plot of particle distribution as a function of $\sin \theta$ and norm $|\bar{\qvec}|$ with Wi, Re=1 at different times (from (a) to (d): t=0.5, 5, 10, 30).}
   \label{hook_KDE_plot}
\end{figure*}

In comparison, Figure \ref{hook_KDE_plot} shows the contour plots of particle distributions for the classical Hookean model with Wi = 1 and Re=1. A "pear" shape is also observed in this case, but the particle elongation is smaller. This distinction is due to the larger entropic forces in the Hookean model when $|\qvec|$ gets larger, prohibiting the particle elongation. As time progresses, the entropic forces and flow effects reach equilibrium. Thus, unlike the Morse and Elastic-plastic models, the particle distribution for the classical Hookean model eventually reaches an equilibrium, not aligned with the flow.

Second, we analyze from the perspective of energy evolution, in order to explain the reduction of stress $\tau_{p21}$. 
When the boundary conditions are taken into consideration, the energy dissipation law can be written as 
\begin{equation}\label{energylaw2}
    \frac{d}{dt} E_{total} = -\Delta + \mathcal{P}_{input}.
\end{equation}
Here, the input energy rate on the boundary yields
\begin{equation}
    \mathcal{P}_{input} = \int_{\Gamma} (\bm \tau \cdot \mathbf{n}) \cdot \uvec \mathrm{d}\x,
\end{equation}
where $\bm \tau = \bm \tau_s + \bm \tau_p$ denotes the total stress tensor, $\mathbf{n}$ is the unit outer normal vector, and $\uvec$ is the velocity of fluid on the boundary. 
%
The total energy $E_{total}$ consists of three parts: kinetic energy, entropy energy and potential energy, which are defined as follows, 
$$
E_{kinetic} = \int_{\Omega} \frac{1}{2} \mbox{Re} |\uvec|^2 \dd \x, 
$$
$$
E_{entropy} = \int_{\Omega} \frac{\epsilon_p}{ \mbox{Wi}} \frac{1}{N_p} \sum_{i=1}^{N_p} \ln \left(\frac{1}{N_p} \sum_{j=1}^{N_p} K_h(\bar{\qvec}_i - \bar{\qvec}_j )\right)  \dd \x, 
$$
and
$$
E_{potential} = 
\int_{\Omega} 
\frac{\epsilon_p}{\mbox{Wi}} \frac{1}{N_p} \sum_{i=1}^{N_p} \bigg{[} \Psi(\bar{\qvec}_i)\bigg{]} \dd \x.
$$

\begin{figure*}
   \includegraphics[width=0.3\linewidth]{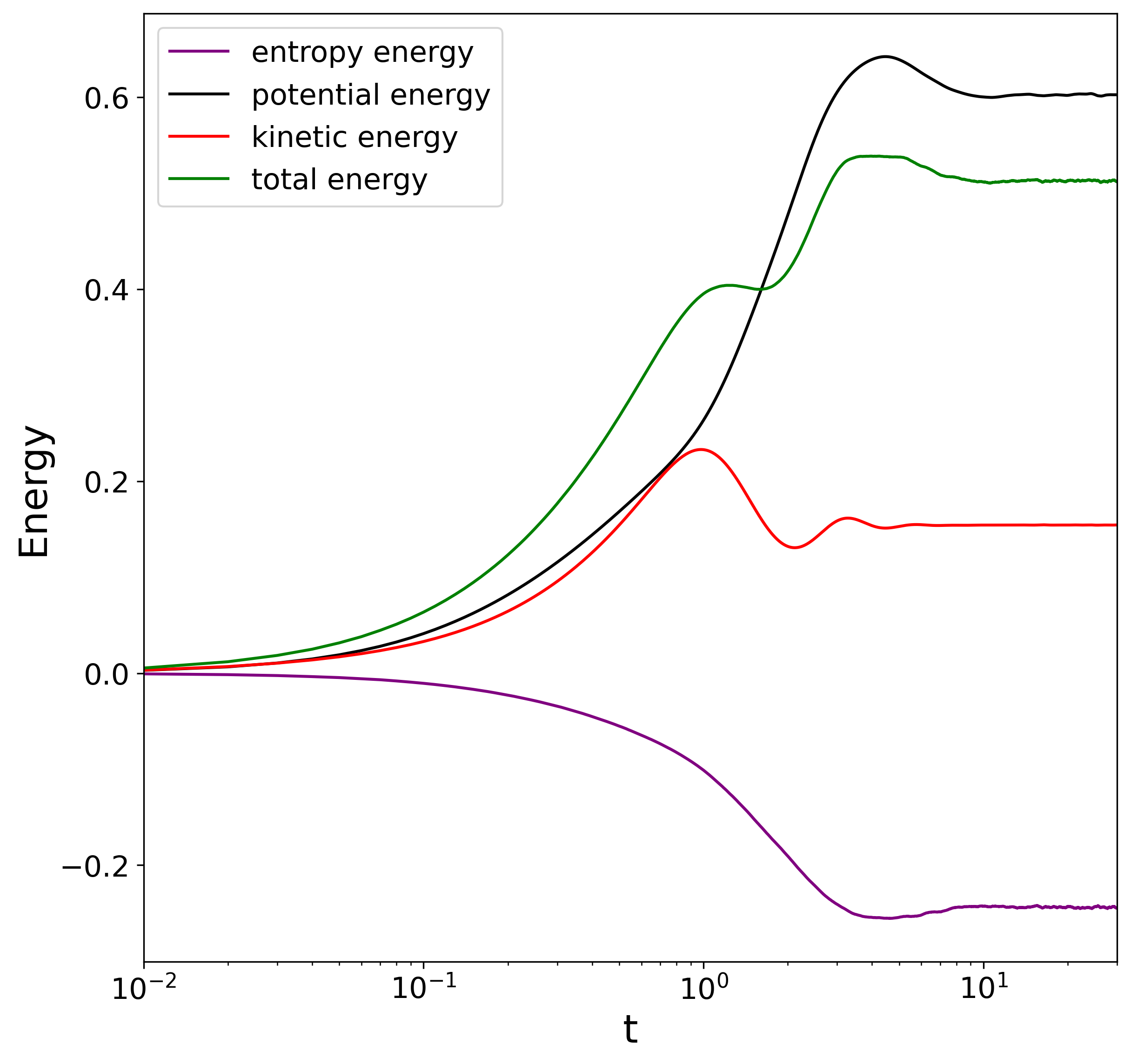}
     \includegraphics[width=0.3\linewidth]{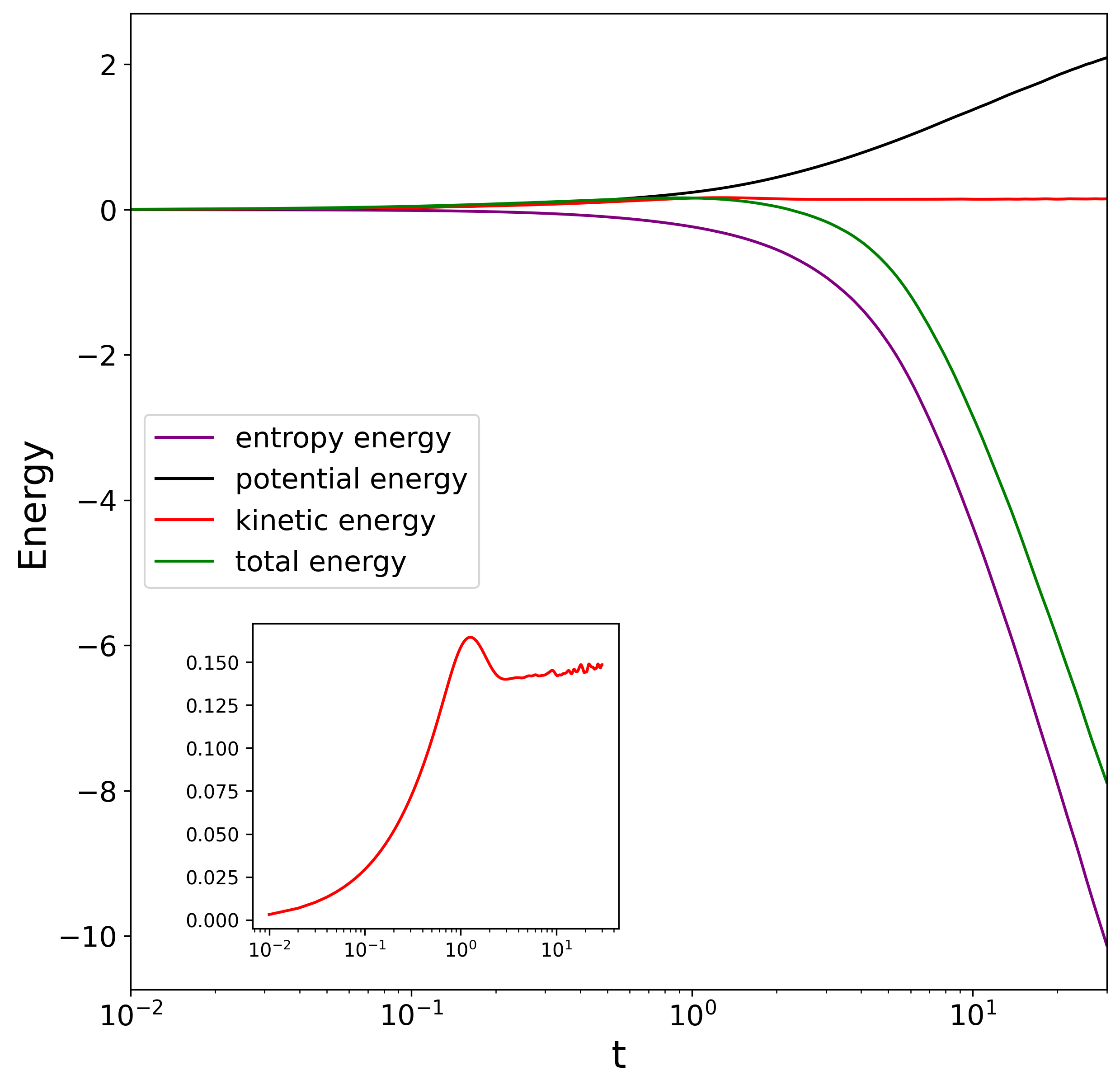}
     \includegraphics[width=0.3\linewidth]{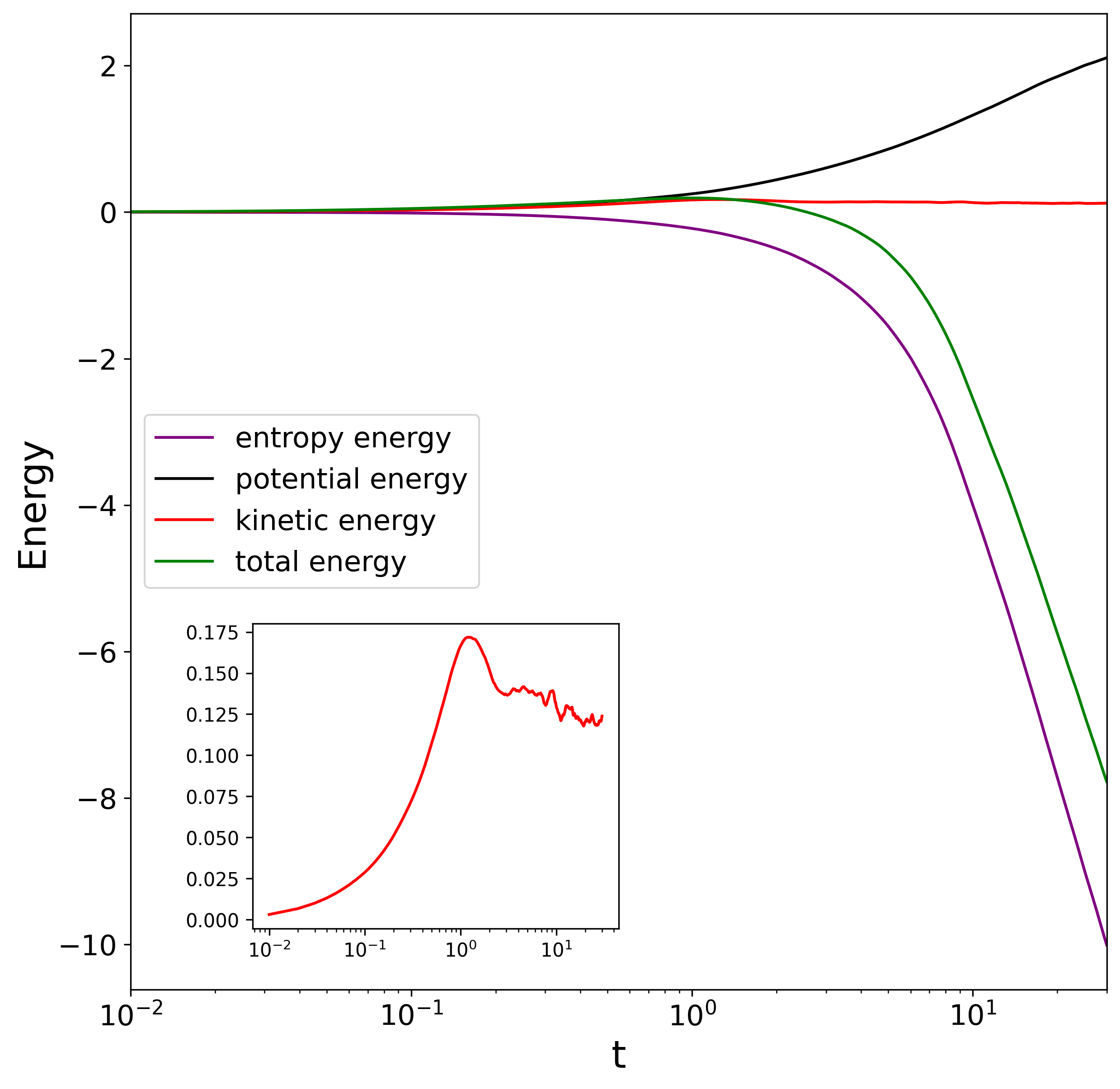}
    \caption{ The time evolution of relative energy of Hookean (left), Morse (middle), and Elastic-plastic model (right) with Wi, Re=1. Time is in a logarithmic scale.}
     \label{energy_constantu}
\end{figure*}

The time evolution of relative energy differences (namely, differences of energy at time $t$ and time $0$) of each type of energy is shown in Figure \ref{energy_constantu}. 
It reveals that for all three models,  the input energy is converted into kinetic energy and potential energy. 
In addition, for the Hookean model (i.e., the left figure in Figure \ref{energy_constantu}), the entropy energy gradually reaches a steady state, since the particle distribution almost reaches equilibrium as time $t$ evolves as in Figure \ref{hook_KDE_plot}. While for Morse and Elastic-plastic models (i.e., the middle and right figures in Figure \ref{energy_constantu}), the entropy energies are decreasing continuously. When the entropy energy is getting smaller, the system is getting more ordered. This is consistent with the previous conclusion that almost all particles are elongated and aligned along the flow for Morse and Elastic-plastic models.

Based on the considerations from the two aspects discussed above, both the Morse and Elastic-plastic models exhibit a trend toward zero stress $\tau_{p21}$. This phenomenon can be attributed to two primary reasons.
Firstly, bond-breaking plays a significant role in this behavior. As time progresses, bond breaking occurs for nearly all particles, leading to the entropic force ($\nabla_{\bar{\qvec}_i}\Psi$) becoming zero for the unbounded particles.
Secondly, the rotation of particles also contributes to the stress reduction phenomenon. Over time, a majority of particles undergo rotation along the x-axis direction. Consequently, $[\nabla_{\bar{\qvec}_i}\Psi]_2$ approaches zero since the head and tail of the rotated particle $\bar{\qvec}_i$ share the same $y$ coordinate.



\vspace{0.5cm}

\subsubsection{Degree of bond breaking and polymer configurations}

We also note that for Morse and Elastic-plastic models, there exist fluctuations along the curves in Figure \ref{uandtaut_plot}, which is related to the bond breaking occurrence. 
%
Actually, the degree of bond breaking can be described by the ratio of broken polymer chains. 
Since particles $\{\bar{\qvec}_i (\x, t) \}_{i=1}^{N_p}$ can be viewed as representative particles, revealing the configuration information of real polymer chains, the breakage ratio can be defined as follows, 
\begin{equation*}
\begin{aligned}
\mbox{Ratio} &= \frac{\mbox{Number of broken polymer chains}}{\mbox{Number of all polymer chains}}\approx \frac{\mbox{Number of particles with $|\bar{\qvec}_i|>Q$ }}{N}.
\end{aligned}
\end{equation*}
In this article, for the Elastic-plastic model, we set $Q=Q_0$ and assume that the bond breaking is irreversible, such that it is counted as broken after the first time it exceeds the threshold. 
For the Morse model, we regard it broken when $|\bar{\qvec}_i|>4=Q$, because the entropic force is negligibly small in that range as shown in Figure \ref{micromacro}. 
 
\begin{figure*}[htpb]
   \centering
\includegraphics[width = 0.9\linewidth]{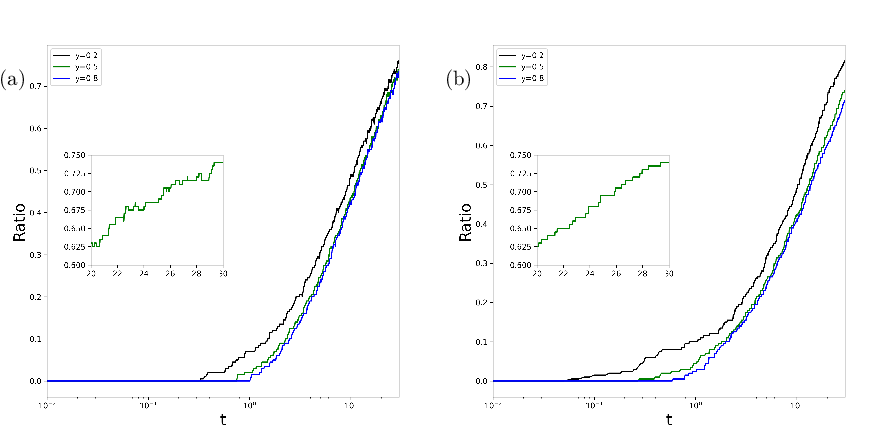}
    \caption{  The time evolution of the breakage ratio for (a) the Morse model and  (b) the Elastic-plastic model, with Wi=1, Re=1 on three locations: $y=0.2$ (black), $y=0.5$ (green), $y=0.8$(blue). Time is in a logarithmic scale. }\label{ratiotplot}
\end{figure*}

Figure \ref{ratiotplot} shows the time evolution of the breakage ratio for Morse and Elastic-plastic models with Wi, Re=1 on different locations. 
For these two models, the ratio is larger on the location $y=0.2$ than that of $y=0.5$ and $0.8$. 
This is because, the closer it is to the moving plane at y=0, the bigger the velocity is and the easier bond breaking occurs. For the Elastic-plastic model, the ratio experiences step-like rises in Figure \ref{ratiotplot}. These steps occur because of certain time interval between every two bond-breaking incidents and the irreversible bond breakage assumption. The fluctuations observed along the curves in Figure \ref{uandtaut_plot} for the Elastic-plastic model can also be attributed to the assumption of irreversible, one-by-one polymer chain breakage. This one-by-one bond breaking leads to fluctuations in the time evolution of velocity, as broken particles tend to approach the velocity of a Newtonian fluid. Additionally, this bond breaking results in the term $\nabla_{\bar{\qvec}_i}\Psi$ becoming zero, and  consequently, the one-by-one bond breaking leads to fluctuations in the shear stress $\tau_{p21}$. 
Regarding the Morse Model, fluctuations are also present along the curves in Figure \ref{uandtaut_plot}, but unlike the Elastic-plastic model, the shape of the curve in Figure \ref{ratiotplot}(a) is irregular rather than the step-like rises, which will be explained through the individual configurations of single particles below. 

\begin{figure*}
   \centering
     \includegraphics[width = 0.9\linewidth]{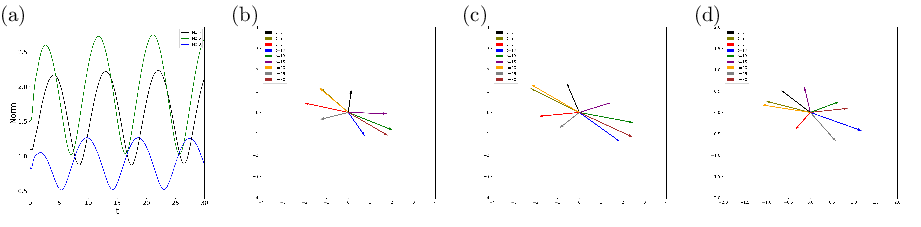}
   \caption{(a) The time evolution of norms of three particles, and (b-d) configurations of one particle at different times respectively for particle No.5, No.6 and No.8; for the Hookean model with Wi, Re=1 on location $y=0.5$.}
   \label{hooknorm568}
\end{figure*}

\begin{figure*}
   \centering
      \includegraphics[width = 0.9\linewidth]{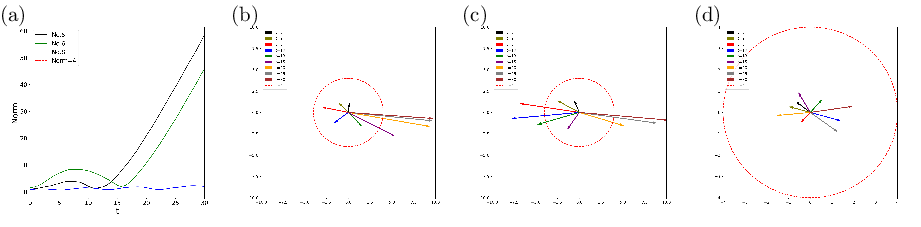}
   \caption{(a) The time evolution of norms of three particles, and (b-d) configurations of one particle at different times respectively for particle No.5, No.6 and No.8; for the Morse model with Wi, Re=1 on location $y=0.5$.}
   \label{morsenorm568}
\end{figure*}

\begin{figure*}
   \centering
     \includegraphics[width = 0.9\linewidth]{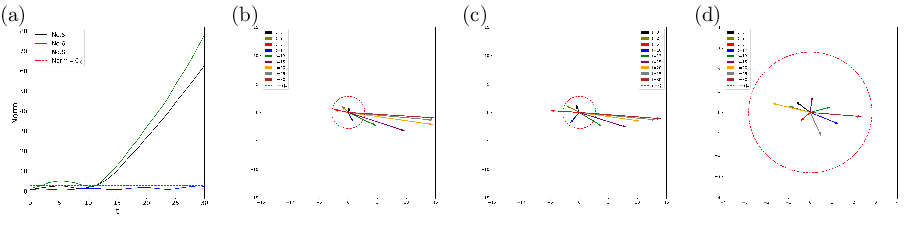}
   \caption{(a) The time evolution of norms of three particles, and (b-d) configurations of one particle at different times respectively for particle No.5, No.6 and No.8; for the Elastic-plastic model with Wi, Re=1 on location $y=0.5$.}   
   \label{plasticnorm568}
\end{figure*}

Next, we provide examples of typical particles, recognizing that individual particle behavior may differ from statistical trends. For the Hookean model, Figure \ref{hooknorm568} illustrates the time evolution of the norms for three different particles and their configurations (from (b) to (d) are particles No.5, No.6, and No.8) at various times. Similarly, Figures \ref{morsenorm568} and \ref{plasticnorm568} show those for the Morse and Elastic-plastic models, with a dashed circle indicating the position of threshold value $Q$ for bond breakage.
In these figures, for the sake of visualization, we have scaled down the extremely long vectors for times when $t>15$. We observe that for all three models, each particle is rotating in the anti-clockwise direction since the lower plane at y=0 is moving. Some particles are elongating and shortening significantly for Morse and Elastic-plastic models, whereas all the 3 particles for the Hookean model keep rotating (the periodic curves in Figure \ref{hooknorm568}(a)) with relatively small magnitude. For the Elastic-plastic model, once the particle norm reaches $Q_0$, irreversible bond breaking occurs and the equation of motion for the particle becomes 
$$
\pp_t \bar{\qvec}_i  - (\nabla_{\x} \uvec)^T\cdot \bar{\qvec}_i =- \frac{1}{2\mathrm{Wi}} \bigg[\bigg(\frac{\sum_{j=1}^{N_p}  \nabla_{\bar{\qvec}_i} K_h(\bar{\qvec}_i- \bar{\qvec}_j)}{\sum_{j = 1}^{N_p} K_h(\bar{\qvec}_i-\bar{\qvec}_j) } + \sum_{k=1}^{N_p} \frac{\nabla_{\bar{\qvec}_i} K_h(\bar{\qvec}_k-\bar{\qvec}_i )}{\sum_{j=1}^{N_p} K_h(\bar{\qvec}_k-\bar{\qvec}_j)}\bigg) + \nabla_{\bar{\qvec}_i} \Psi(\bar{\qvec}_i) \bigg]
$$
with $\nabla_{\bar{\qvec}_i} \Psi(\bar{\qvec}_i) = 0$. In this case, the term $(\nabla_{\x} \uvec)^T\cdot \bar{\qvec}_i$ is dominant, and the particle is rotated until $[\bar{\qvec}_i]_2=0$, as shown in Figure \ref{plasticnorm568} (b) and (c). For the Morse model, since there is no explicit abrupt bond breaking event in the Morse potential, the particle could be elongated and then shortened and finally elongated very long continuously as in Figure \ref{morsenorm568}(c). In the end, the particle is rotated along the macroscopic flow and stays at broken state. The elongation and shortening of particles result in the irregular curve of Ratio in Figure \ref{ratiotplot}(a).
Furthermore, we note that particle No.6 rotates faster for the Elastic-plastic model than the Morse model, as shown in Figure \ref{morsenorm568} (c) and \ref{plasticnorm568} (c). The reason is that once the particle is elongated longer than $Q_0$, the term $\nabla_{\bar{\qvec}_i} \Psi(\bar{\qvec}_i)$ remains zero for the Elastic-plastic model;  while for the Morse model, the term $\nabla_{\bar{\qvec}_i} \Psi(\bar{\qvec}_i)$ can be large if the particle is shortened, which counterbalances the effect of $- (\nabla_{\x} \uvec)^T\cdot \bar{\qvec}_i$.

Then we give two possible explanations for the smooth fluctuations in the time evolutions of velocity and shear stress for the Morse model in Figure \ref{uandtaut_plot}. 
One is based on the definition of shear stress
$$
\tau_{p21} = \frac{\epsilon_p}{\mathrm{Wi}} \frac{1}{N_p} \sum_{i=1}^{N_p} \frac{1}{\beta^2}(1-e^{-\beta|\bar{\qvec}_{i}|-\alpha|\bar{\qvec}_{i}|^3})e^{-\beta|\bar{\qvec}_{i}|-\alpha|\bar{\qvec}_{i}|^3}(\beta |\bar{\qvec}_{i}|^{-1} + 3\alpha |\bar{\qvec}_{i}|)  [\bar{\qvec}_{i}]_2 [\bar{\qvec}_{i}]_1.
$$
With irregular but continuous changes in particle norms, 
we obtain fluctuations in the nonlinear coefficient before the term $[\bar{\qvec}_{i}]_2 [\bar{\qvec}_{i}]_1$ inside the summation. When the coefficient from some particular particle experience continuous changes, in the average sense (summation), the stress shows small and smooth fluctuations. 
In another explanation, the Morse potential 
$$
\Psi(\mathbf{q})=\frac{1}{2\beta^2} \bigg(1-e^{-\beta|\mathbf{q}|-\alpha |\mathbf{q}|^3 }\bigg)^2
$$
can be viewed as a nonlinear spring potential.
It is known that the nonlinear spring potential, such as a polynomial potential of degree greater than two, has a variable frequency. The effects of fluid on the "spring" can be regarded as an external force. When the frequency of the external force coincides with the frequency of the "spring", resonance could occur. Thus, the small smooth fluctuations could correspond to certain large frequency or resonance behaviors.

While for the Hookean model, although there exist changes in the norm and orientations for individual particles, there are no noticeable fluctuations in the time evolution of the shear stress in Figure \ref{uandtaut_plot}. The reason may be that $\tau_{p21} = \frac{\epsilon_p}{\mathrm{Wi}} \frac{1}{N_p} \sum_{i=1}^{N_p}[\bar{\qvec}_{i}]_2 [\bar{\qvec}_{i}]_1$ is linear with respect to $[\bar{\qvec}_{i}]_2 [\bar{\qvec}_{i}]_1$, and the initial set of particles has random distribution. Furthermore, from  Figure \ref{hooknorm568}, we notice that the norm of each particle changes periodically, i.e., the particle is elongated and shortened periodically as it rotates. The randomness is somehow preserved and then there is only a smooth transition without fluctuations in the average (statistical) sense due to the summation formula.   


\vspace{0.5cm}

\subsubsection{Rheology of the models}

In this section, we focus on the rheology of the three different models. Specifically, we examine the impact of shear rate through variations in the bottom plate velocity, which represents the characteristic velocity $\tilde{U}$. By keeping other characteristic parameters constant, an increase in shear rate corresponds to a simultaneous increase in both the Reynolds number Re and the Weissenberg number Wi.

 
%


\begin{figure*}
   \centering
     \includegraphics[width = 0.9\linewidth]{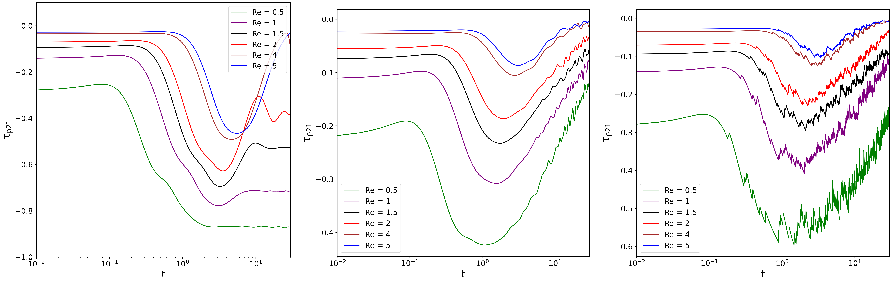}
   \caption{ Hookean (left), Morse (middle) and Elastic-plastic model (right): The time evolution of shear stress on location $y=0.5$ with different Wi and Re (Wi=Re in these cases). Time is in a logarithmic scale.} 
   \label{morse_shearstress_t}
\end{figure*}


\begin{figure*}
   \centering
   \includegraphics[width=0.9\linewidth]{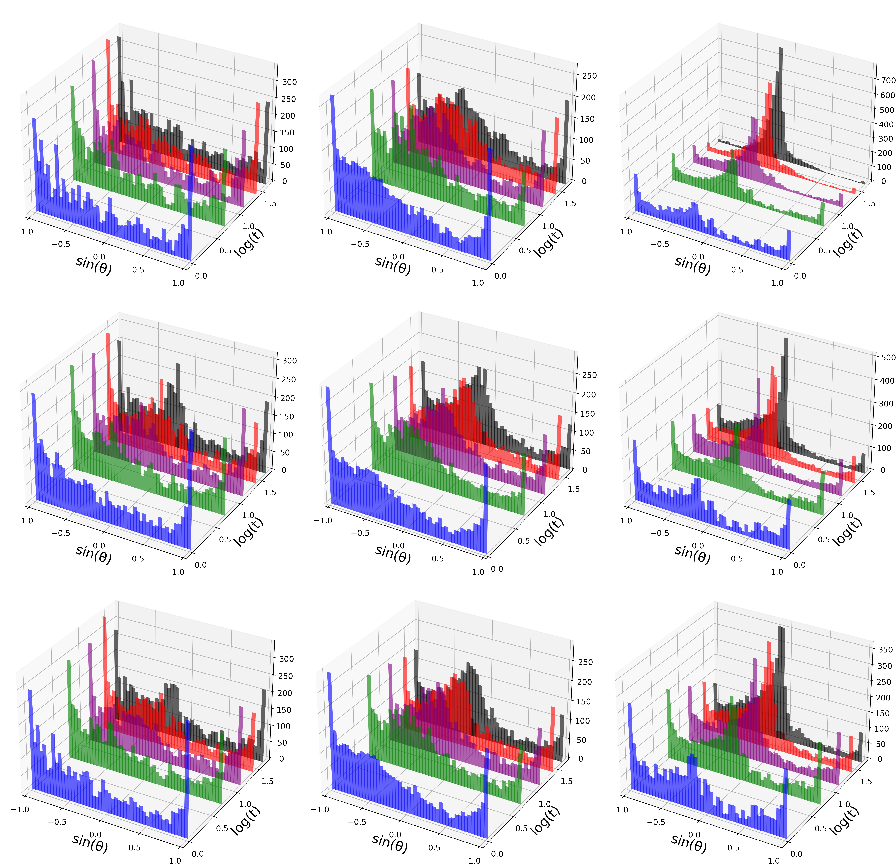}
   \caption{3D bar plot as a function of $\sin \theta$ at all locations with Wi, Re = 0.5 (left), Wi, Re=1 (middle) and Wi, Re=5 (right) at different times ($t=1, 2, 3, 4, 5$). First row: Hookean model; Second row: Morse model; Third row: Elastic-plastic model.}
   \label{3Dbarplot_ally_diffWe}
\end{figure*}

\begin{figure}
   \centering 
    \includegraphics[width=0.9\linewidth]{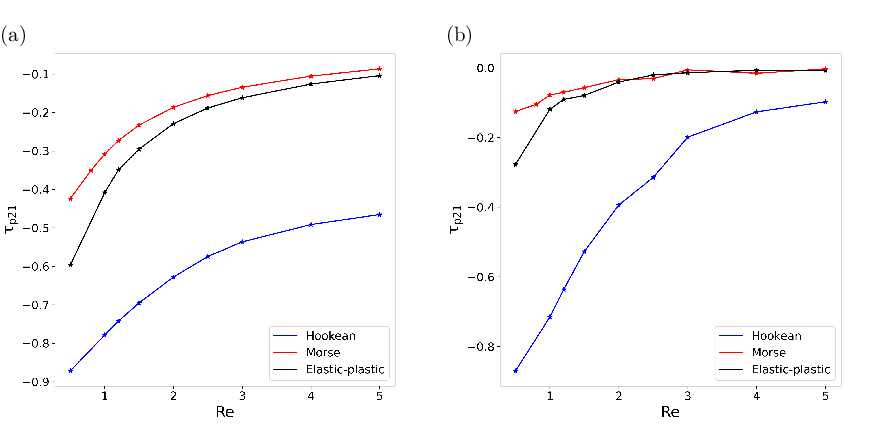}
   \caption{ (a) The minimum value (maximum magnitude) of shear stress $\tau_{p21}$ during the evolution;  and (b) the steady-state value of $\tau_{p21}$, as a function of Re on location $y=0.5$. } 
   \label{tau_Re_constantu}
\end{figure}

Figure \ref{morse_shearstress_t}  shows the time evolution of stress $\tau_{p21}$ with different shear rates (namely, different Re and Wi) for the Hookean, Morse, and Elastic-plastic models. 
Since numerical results show similar behaviors on different locations, Figure \ref{morse_shearstress_t}  only shows the time evolutions at location $y=0.5$ for illustration.  It indicates that stress $\tau_{p21}$ has a smaller minimum (less negative) and approaches zero more quickly with the larger shear rates (larger Re and Wi) for all three models. It could be explained by statistical analysis of particle configurations. Figure \ref{3Dbarplot_ally_diffWe} shows the 3D bar plot of particle configurations as a function of  $\sin \theta$ for Hookean, Morse, and Elastic-plastic models with different Wi and Re at different times ($t=1, 2, 3, 4, 5$). 
The z-axis represents the number of particles in the whole domain. 
The 3D bar plots show that the polymer chains tend to rotate earlier to align with the flow when the Wi and Re are larger. 
For instance, for the Morse model, with Wi, Re=1, there are significant number of particles near $\sin \theta=0$ at time $t=5$, while with Wi, Re=5, there are already significant number of particles near $\sin \theta=0$ at time $t=2$. As $\sin \theta=0$ implies $[\nabla_{\bar{\qvec}_i}\Psi]_2 =0$ for the particle $\bar{\qvec}_i$, more number of particles close to $\sin \theta=0$ indicate that fewer particles contribute to stress $\tau_{p21}$ with bigger Wi and Re, which leads to a smaller minimum and approaches zero more quickly.

%
 


\begin{figure*}[htpb]
   \centering
    \includegraphics[width=0.9\linewidth]{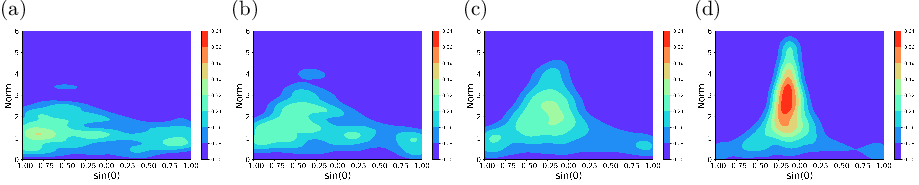}
   \caption{Hookean model: Contour plot of particle distribution as a function of $\sin(\theta)$ and norm with different Wi and Re at the steady state (from (a) to (d): Wi, Re=0.5; Wi, Re=1; Wi, Re=2 and Wi, Re=4).}
   \label{hook_KDE_plot_diffWe}
\end{figure*}

%
Figure\ref{tau_Re_constantu} shows
the minimum value (maximum magnitude) of shear stress $\tau_{p21}$ during the evolution and the steady-state value of $\tau_{p21}$, as a function of the Reynolds number Re. 
The curves suggest the shear-thinning behavior of the three models, which is consistent with Figure \ref{morse_shearstress_t}. 
At the steady state, 
for Morse and Elastic-plastic models, the shear stresses remain near zero as Re varies, which is consistent with the contour and  3D bar plots of particle configurations. 
For the shear-thinning behavior of the Hookean model, the first row in Figure \ref{3Dbarplot_ally_diffWe} illustrates that with larger Re, more polymers align with the flow direction (close to $\sin(\theta)=0$). Moreover, Figure \ref{hook_KDE_plot_diffWe} presents the contour plots of particle distribution as a function of $\sin(\theta)$ and norm at the steady state. It also shows that the "pear" shape is more apparent and vertical around $\sin(\theta)=0$ as Wi and Re get larger, indicating that more particles align with the flow direction at the steady state. 
As a result, fewer particles contribute to stress $\tau_{p21}$ with larger Re, leading to smaller magnitude for $\tau_{p21}$ in the shear-thinning behavior for the Hookean model.




\vspace{0.5cm}

\subsection{Simple shear flow with periodic $U$}

In the previous section, when the velocity of the moving plane is constant, for Morse and Elastic-plastic models, almost all particles eventually elongate in one direction. 
In order to investigate more complicated phenomenon, in this section, we consider the shear flow case where the lower plane move in the $x$ direction with a periodic velocity $U$ with respect to time $t$:
$$
U = A \sin(\omega t).
$$
Here, we set $A = 1$, $\omega= \pi/4$ and the corresponding period is $T = 2\pi/\omega = 8$.  
Other parameters are set as those in the last section. The periodic velocity of the moving plane will lead to the periodic fluid velocity for the whole domain, and thus, the distances between polymer beads will also vary. For the Elastic-plastic model, the assumption of irreversible bond breaking is also adopted. 


%
%

\subsubsection{Velocity and shear stress evolution with periodic $U$}

\begin{figure*}
   \centering
    \includegraphics[width=0.9\linewidth]{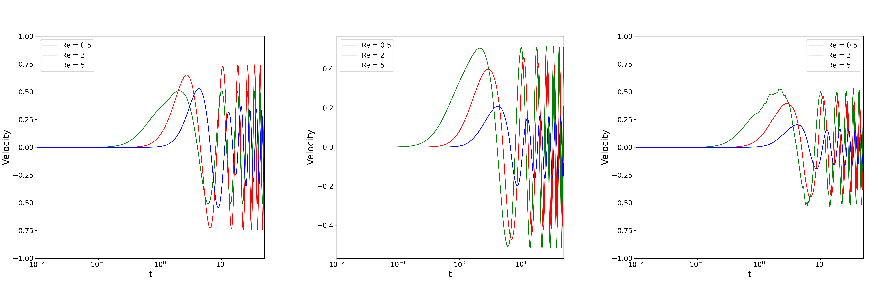}
   \caption{   Hookean (left), Morse (middle) and Elastic-plastic model (right) in the periodic $U$ case: The time evolution of velocity at location $y=0.5$ with different Wi and Re (Wi=Re in these cases). Time is in a logarithmic scale.}  
   \label{periodicu_tplot}
\end{figure*}
\begin{figure*}
   \centering
   \includegraphics[width=0.9\linewidth]{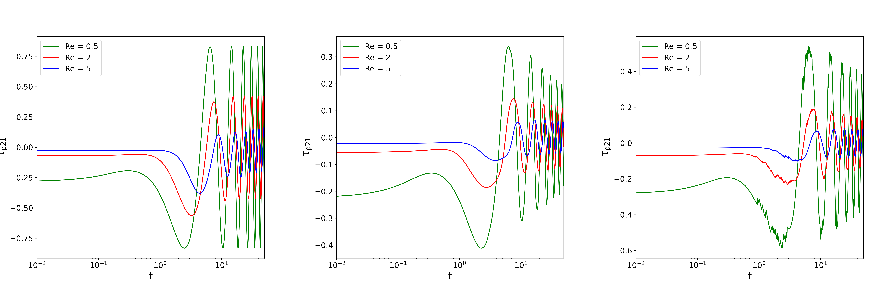}
  \caption{  Hookean (left), Morse (middle) and Elastic-plastic model (right) in the periodic $U$ case: The time evolution of shear stress at location $y=0.5$ with different Wi and Re. Time is in a logarithmic scale. }  
   \label{periodic_shearstress_tplot}
\end{figure*}
\begin{figure*}[htpb]
   \centering
   \begin{overpic}[width = 0.45\linewidth]{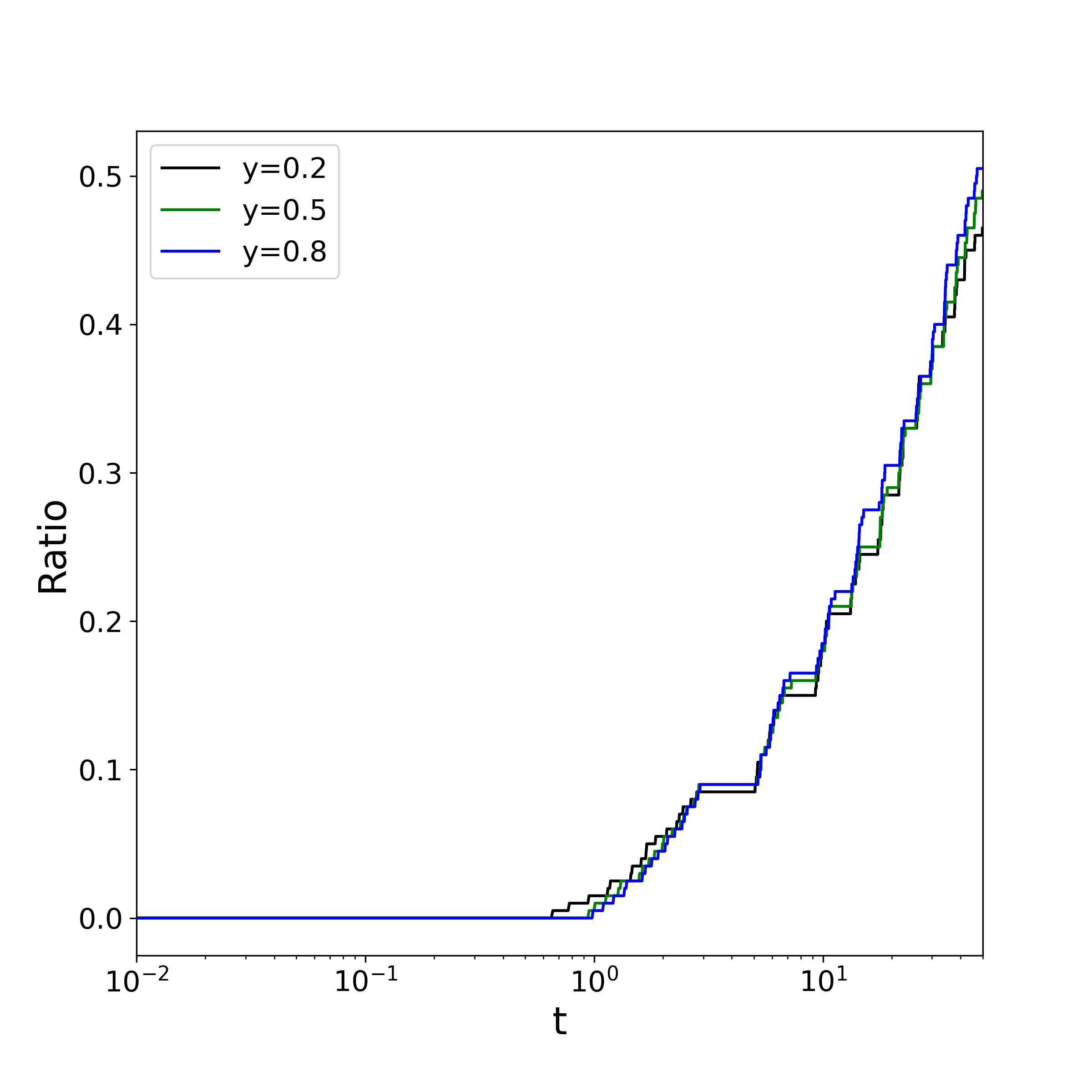}
   \end{overpic}
    \caption{  The time evolution of Ratio for the Elastic-plastic model in the periodic $U$ case with Wi, Re =0.5 on different locations: $y=0.2$ (black), $y=0.5$ (green), $y=0.8$(blue). Time is in a logarithmic scale.}  
    \label{periodic_ratio_tplot}
\end{figure*}

Figures \ref{periodicu_tplot} and \ref{periodic_shearstress_tplot} show the time evolution of velocity and shear stress $\tau_{p21}$ with different Wi and Re for the three models. Notice that the velocities and shear stresses show periodic behaviors due to the periodic oscillation of U at bottom plane. 
%
%
%
Similar to the previous section, there exist non-smooth fluctuations in velocity and shear stress for the Elastic-plastic model, which  is due to bond breaking. The bond breaking is reflected in Figure \ref{periodic_ratio_tplot}, which presents the time evolutions of the breakage ratio for the Elastic-plastic model in the case of periodic $U$ with Wi, Re =0.5 on different locations. Notice that the ratio also rises in the form of steps because of the irreversible assumption of bond breaking for the Elastic-plastic model. Figure \ref{periodic_shearstress_tplot} also shows that, for Morse and Elastic-plastic models, the amplitudes of the stress $\tau_{p21}$ are getting smaller as time evolves, which is different from the constant magnitude in the Hookean model. As before, we try to explain these phenomena from two aspects: the distribution of particles and the energy evolution.

\begin{figure*}[htpb]
   \centering
  \includegraphics[width=0.9\linewidth]{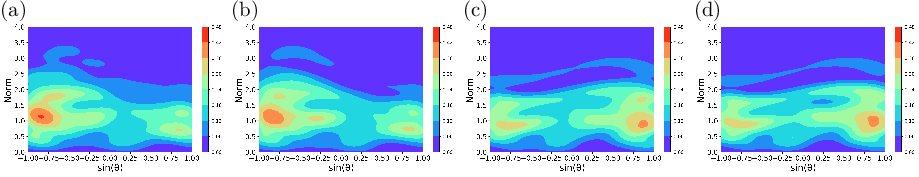}
   \caption{   Hookean model in the periodic $U$ case: Contour plots of particle distribution as a function of $\sin \theta$ and norm with Wi, Re=0.5 at different times (from (a) to (d): t=2, 10, 24, 48). }  
   \label{hookperiodicKDE}
\end{figure*}

\begin{figure*}[htpb]
   \centering
  %
   \includegraphics[width=0.9\linewidth]{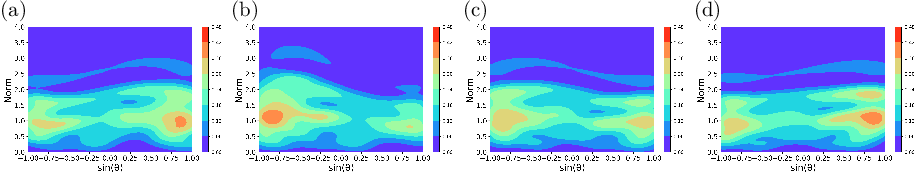}
   \caption{  Hookean model in the periodic $U$ case: Contour plots of particle distribution as a function of $\sin \theta$ and norm with Wi, Re = 0.5 at different times of one period (from (a) to (d): t=40, 42, 44, 45). }  
   \label{hookperiodicKDE2}
\end{figure*}

\begin{figure*}[htpb]
   \centering
    \includegraphics[width=0.9\linewidth]{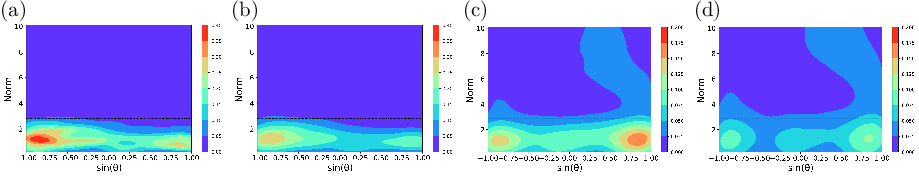}
   \caption{  Elastic-plastic model in the periodic $U$ case: Contour plots of particle distribution as a function of $\sin \theta$ and norm with Wi, Re = 0.5 at different times of one period (from (a) to (d): t=2, 10, 24, 48). }  
   \label{plasticperiodicKDE}
\end{figure*}

\begin{figure*}[htpb]
   \centering
  \includegraphics[width=0.9\linewidth]{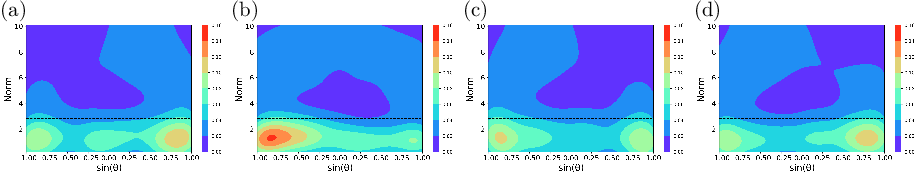}
   \caption{  Elastic-plastic model in the periodic $U$ case: Contour plots of particle distribution as a function of $\sin \theta$ and norm with Wi, Re = 0.5 at different times of one period (from (a) to (d): t=40, 42, 44, 45). }  
   \label{plasticperiodicKDE2}
\end{figure*}

\begin{figure*}[htpb]
   \centering
   \includegraphics[width=0.9\linewidth]{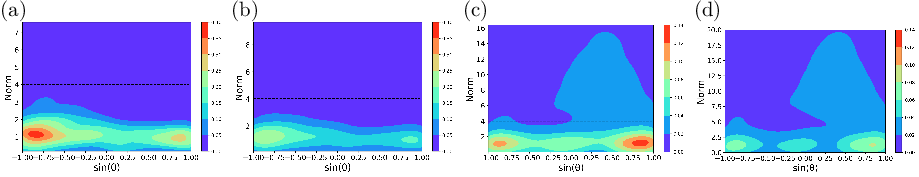}
   \caption{   Morse model in the periodic $U$ case: Contour plots of particle distribution as a function of $\sin \theta$ and norm with Wi, Re=0.5 at different times (from (a) to (d): t=2, 10, 24, 48). }  
   \label{morseperiodicKDE}
\end{figure*}

\begin{figure*}[htpb]
   \centering
   \begin{overpic}[width=0.23\linewidth]{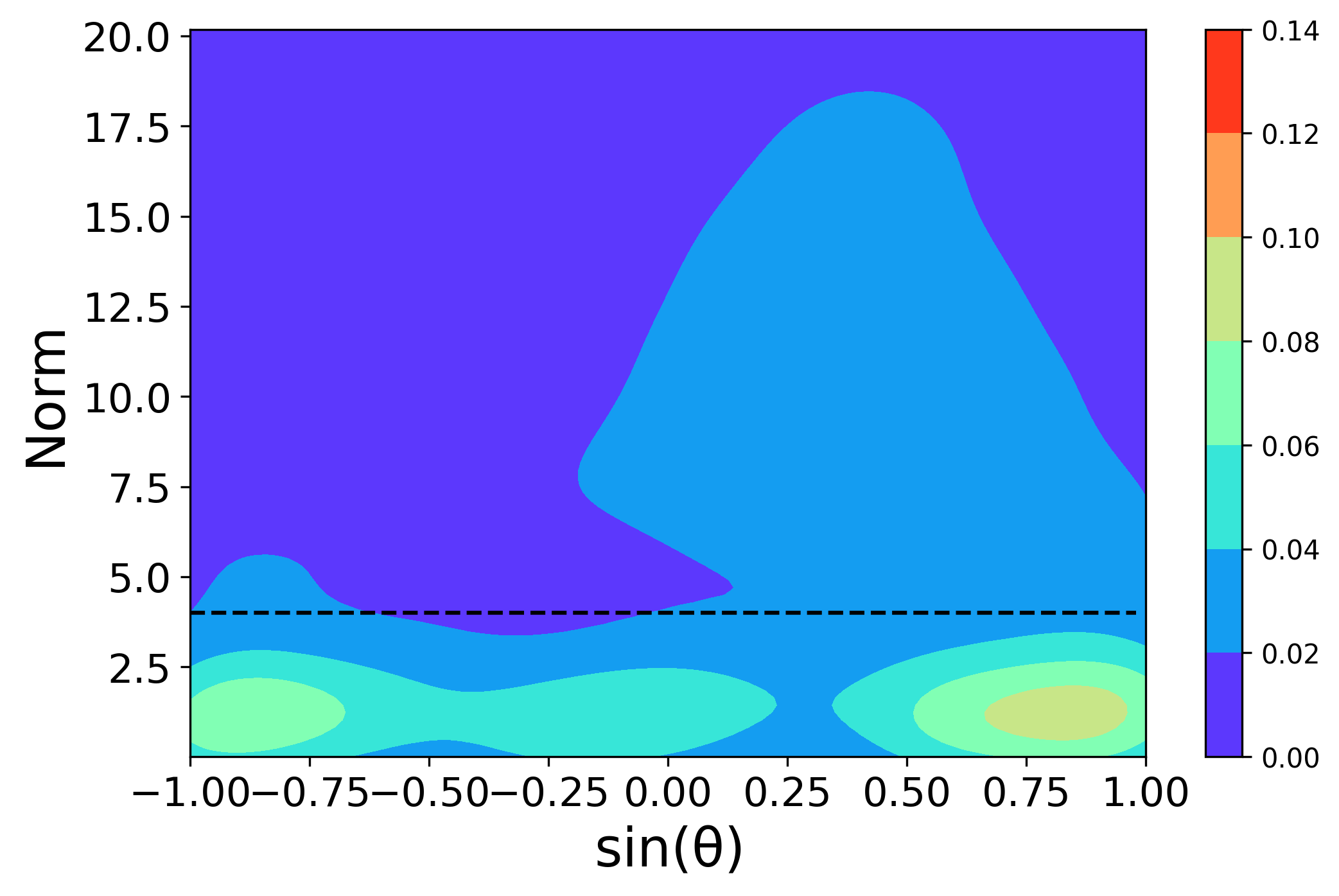}
   \put(0, 71){(a)}
   \end{overpic}
   \begin{overpic}[width=0.23\linewidth]{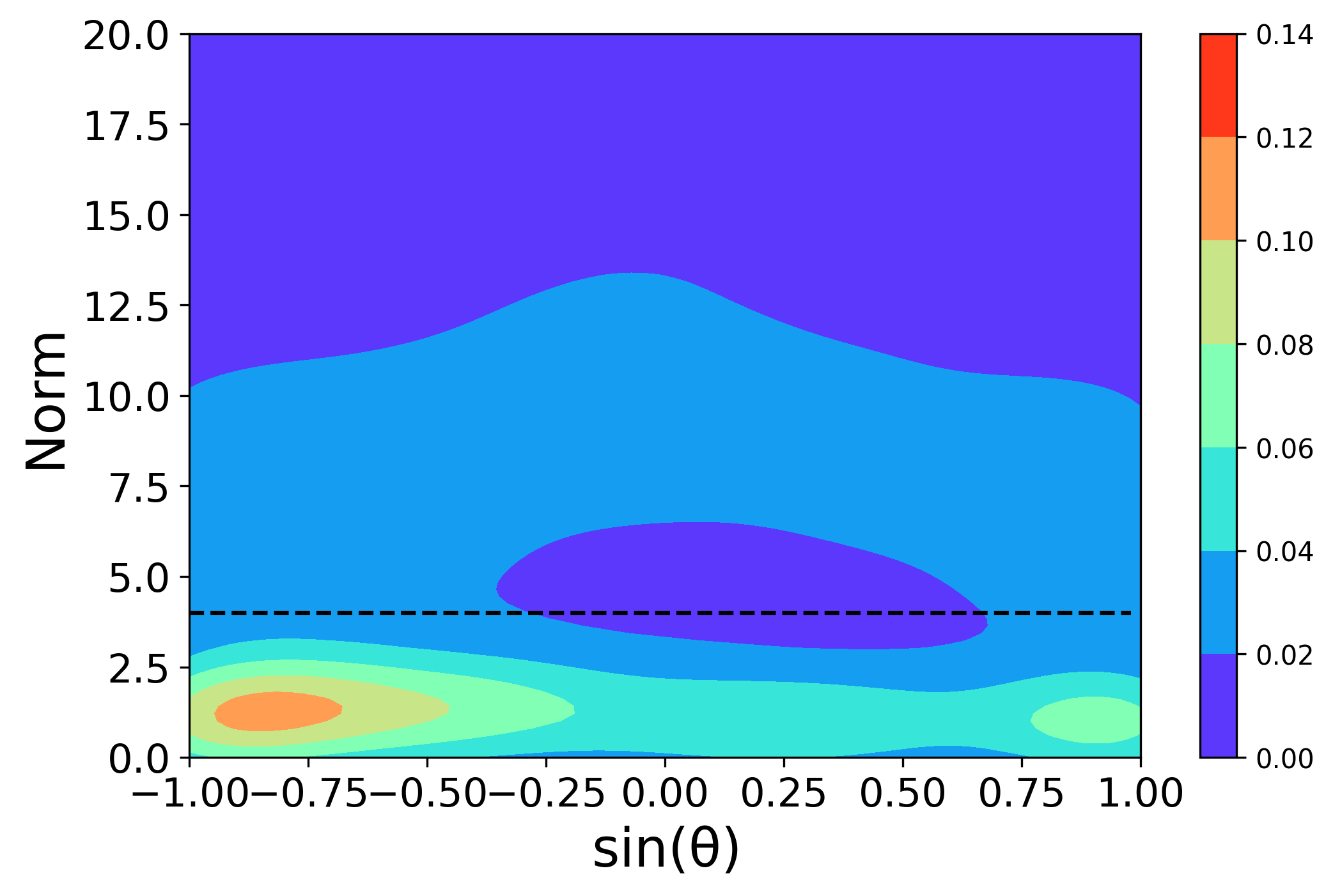}
     \put(0, 71){(b)}
   \end{overpic}
   \begin{overpic}[width=0.23\linewidth]{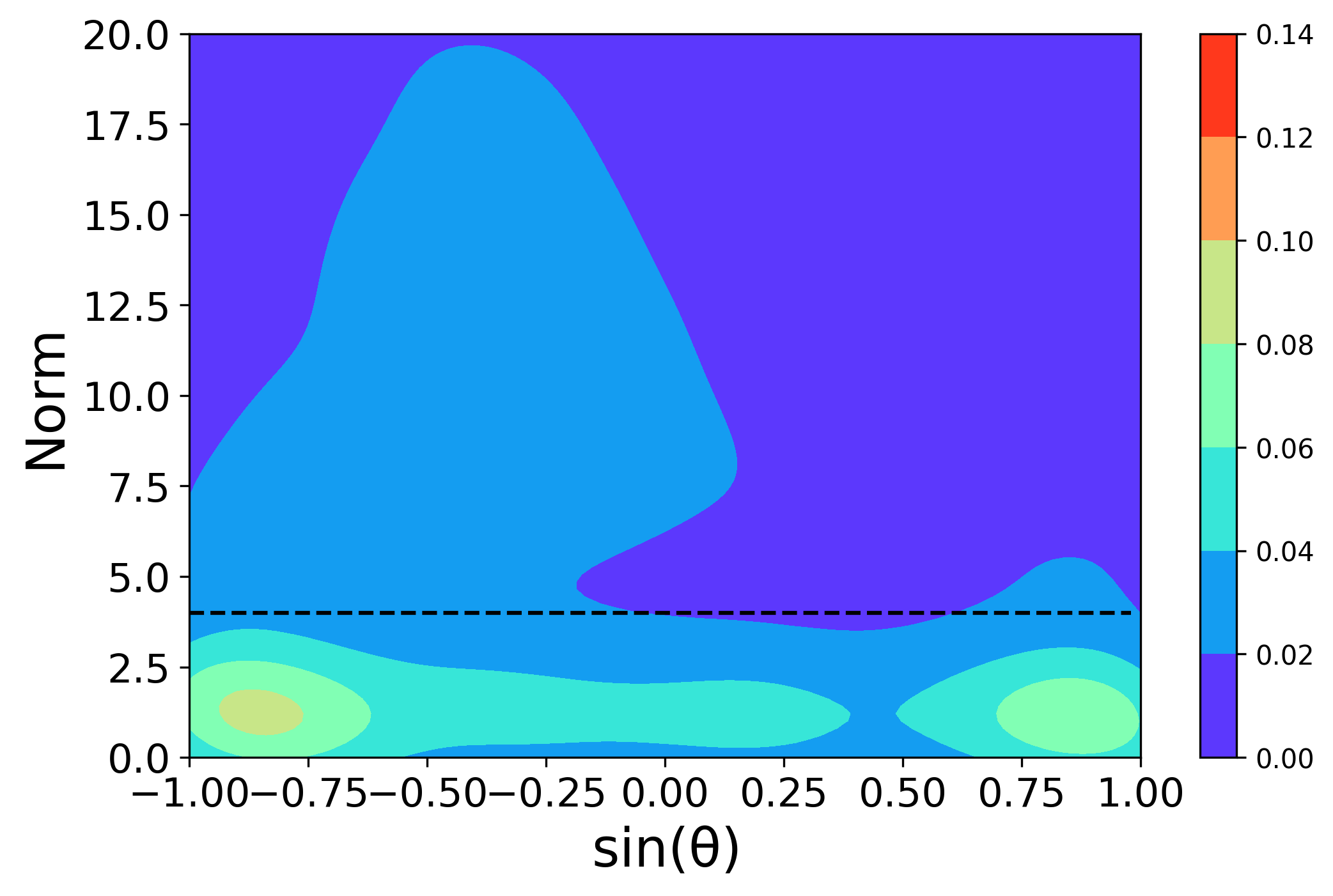}
      \put(0, 71){(c)}
   \end{overpic}
    \begin{overpic}[width=0.23\linewidth]{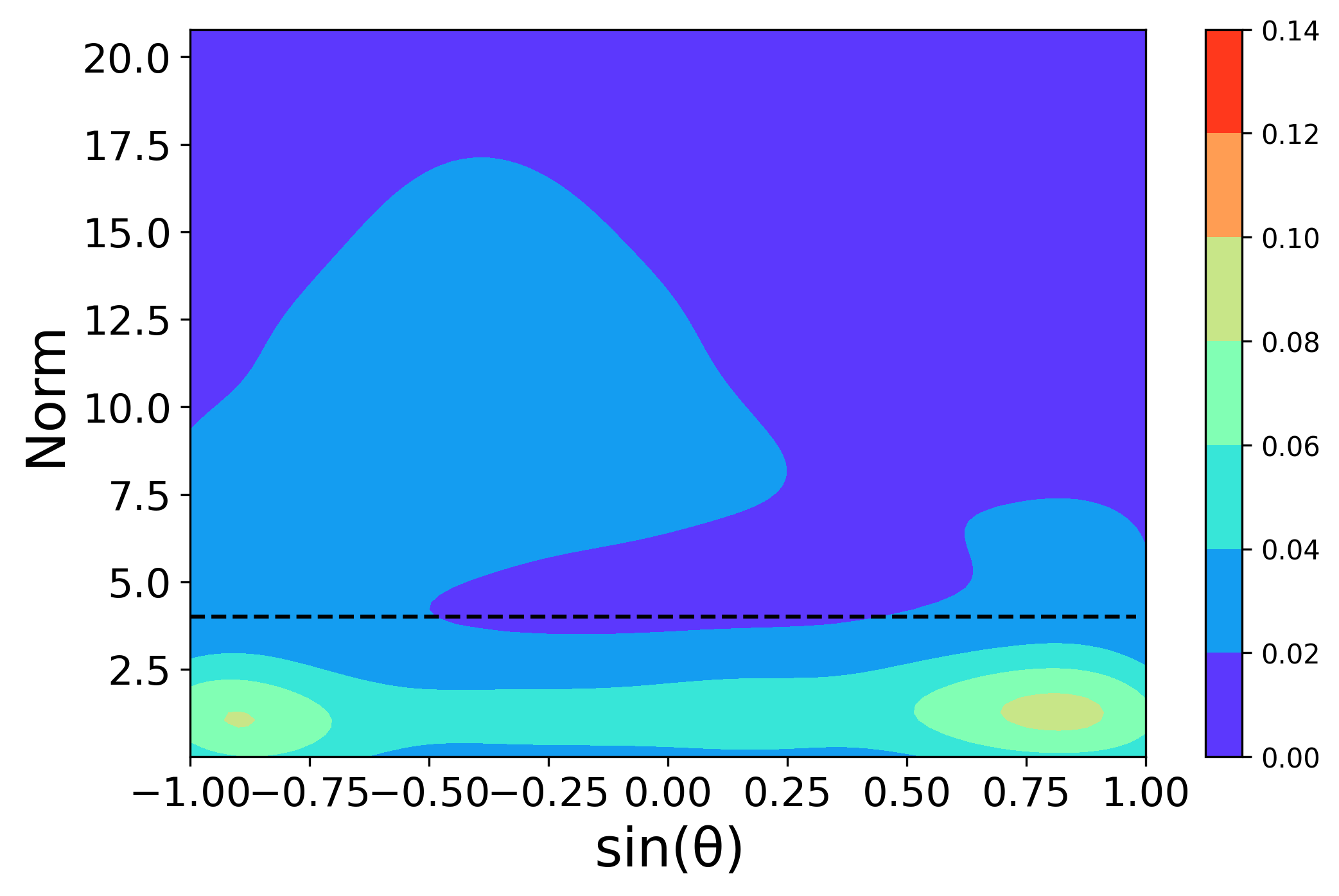}
   \put(0, 71){(d)}
  \end{overpic}
   \caption{ Morse model in the periodic $U$ case: Contour plots of particle distribution as a function of $\sin \theta$ and norm with Wi, Re = 0.5 at different times of one period (from (a) to (d): t=40, 42, 44, 45). }  
   \label{morseperiodicKDE2}
\end{figure*}

The contour plots of particle distribution as a function of $\sin \theta$ and norm with Wi, Re=0.5 at different times for the three models are presented in Figures \ref{hookperiodicKDE}, \ref{hookperiodicKDE2}, Figures \ref{plasticperiodicKDE}, \ref{plasticperiodicKDE2}, and Figures \ref{morseperiodicKDE}, \ref{morseperiodicKDE2}, respectively.  We first emphasize that the periodic $U$ leads to similar periodic behaviors of particle distribution  with period $T=8$, see Figure \ref{hookperiodicKDE} (d) and Figure \ref{hookperiodicKDE2} (a) for the Hookean model, Figure \ref{plasticperiodicKDE} (d) and Figure \ref{plasticperiodicKDE2} (a) for the Elastic-plastic model and Figure \ref{morseperiodicKDE} (d) and Figure \ref{morseperiodicKDE2} (a) for the Morse model. 
Figures \ref{hookperiodicKDE} and \ref{hookperiodicKDE2} also show that there are no significant changes in the particle distributions for the Hookean model as time goes on.  For Morse and Elastic-plastic models in Figures \ref{plasticperiodicKDE}-\ref{morseperiodicKDE2}, as time evolves, many particles are elongated to a relatively large magnitude, which is consistent with Figure \ref{periodic_ratio_tplot} that bond breaking occurs for almost half of the particles. 
Moreover, different from the previous case with constant U, when time $t$ is large, the contour plots of particle distribution with periodic U don't show a "pear", "stripe" or line shape. Actually, as shown in Figures \ref{plasticperiodicKDE2} and \ref{morseperiodicKDE2}, 
$\sin \theta$ tends to be $\pm 1$ in the whole period for large $t$, which means that many remaining unbroken particles tend to be perpendicular to the moving plane.

\begin{figure*}
   \centering
     \includegraphics[width=0.9\linewidth]{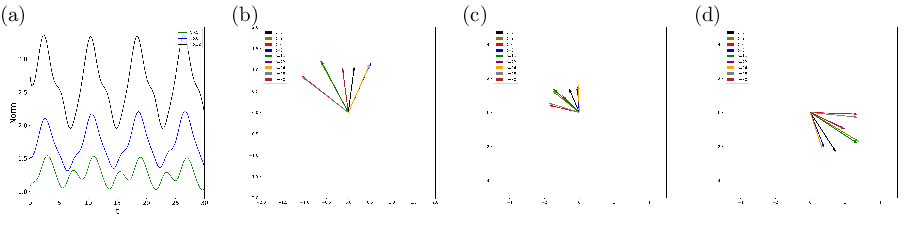}
   \caption{ Hookean model in the periodic $U$ case: (a) Time evolution of norms of three different particles,  and (b-d) configurations of the three particles (particles No.5, No.6, and No. 12 from (b) to (d)), at different times 
  with Wi, Re=0.5 on location $y=0.5$.  }
   \label{hooknorm568periodu}
\end{figure*}

\begin{figure*}
   \centering
     \includegraphics[width=0.9\linewidth]{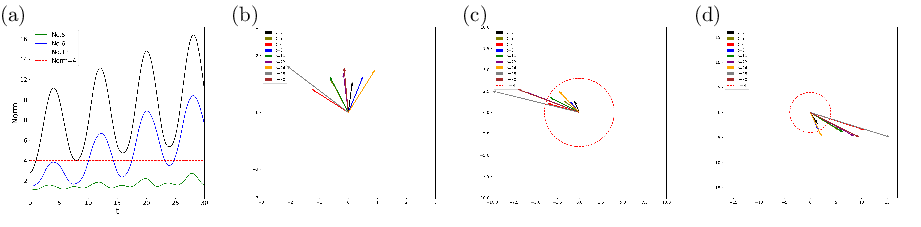}
   \caption{ Morse model in the periodic $U$ case: (a) Time evolution of norms of three different particles,  and (b-d) configurations of the three particles (particles No.5, No.6, and No. 12 from (b) to (d)), at different times 
  with Wi, Re=0.5 on location $y=0.5$.  }
   \label{morsenorm568periodu}
\end{figure*}

\begin{figure*}
   \centering
    \includegraphics[width=0.9\linewidth]{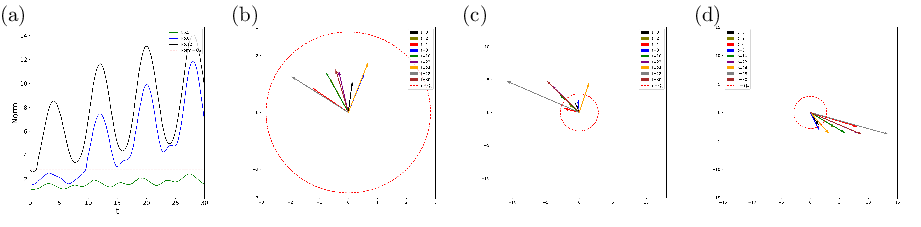}
   \caption{Elastic-plastic model in the periodic $U$ case: (a) Time evolution of norms of three different particles,  and (b-d) configurations of the three particles (particles No.5, No.6, and No. 12 from (b) to (d)), at different times with Wi, Re=0.5 on location $y=0.5$. The red dahsed line and circle represent the threshold for bond breakage.} 
   \label{plasticnorm568periodu}
\end{figure*}

Now, we investigate the configurations of some typical individual particles. Since the contour plots of particles present the overall configurations of particles, 
the configuration and behavior of selected individual particles may be different from the statistical phenomenon. 
Figures \ref{hooknorm568periodu},  \ref{morsenorm568periodu} and \ref{plasticnorm568periodu} plot the time evolutions of norms of three different particles and their configurations at different times with Wi, Re=0.5 on location $y=0.5$, respectively for Hookean, Morse, and Elastic-plastic models. 
For the Hookean model, the configurations of the particles show periodic behaviors with period $T=8$ for both the orientation and norm. 
For Morse and Elastic-plastic models, the orientation of each particle shows similar periodic behaviors with period $T=8$, but at the same time, the norm (magnitude) of each particle tends to increase as time evolves, especially for particles with larger initial norms. 
This phenomenon is consistent with the contour plots in Figures \ref{plasticperiodicKDE} and \ref{morseperiodicKDE}. 
Since the equation of motion for each particle reads as follows,  
$$
\pp_t \bar{\qvec}_i  - (\nabla_{\x} \uvec)^T\cdot \bar{\qvec}_i =- \frac{1}{2\mathrm{Wi}} \bigg[\bigg(\frac{\sum_{j=1}^{N_p}  \nabla_{\bar{\qvec}_i} K_h(\bar{\qvec}_i- \bar{\qvec}_j)}{\sum_{j = 1}^{N_p} K_h(\bar{\qvec}_i-\bar{\qvec}_j) } + \sum_{k=1}^{N_p} \frac{\nabla_{\bar{\qvec}_i} K_h(\bar{\qvec}_k-\bar{\qvec}_i )}{\sum_{j=1}^{N_p} K_h(\bar{\qvec}_k-\bar{\qvec}_j)}\bigg) + \nabla_{\bar{\qvec}_i} \Psi(\bar{\qvec}_i) \bigg]. 
$$
When the particle is elongated, the term $\nabla_{\bar{\qvec}_i} \Psi(\bar{\qvec}_i)$ is small and even equals to zero when $|\bar{\qvec}| > Q_0$ for the Elastic-plastic model.  In this case, the term $- (\nabla_{\x} \uvec)^T\cdot \bar{\qvec}_i$ dominates the dynamics, and thus the particle is further elongated periodically since $\nabla_{\x} \uvec$ is periodic.

%


%

\begin{figure*}
    \includegraphics[width=0.9\linewidth]{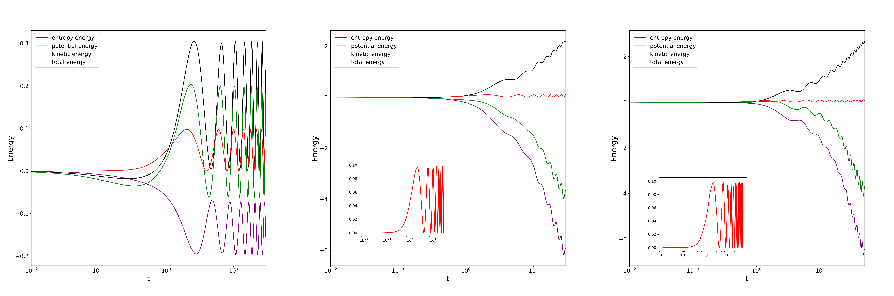}
    \caption{   Hookean (left), Morse (middle) and Elastic-plastic model (right) in the periodic $U$ case: The time evolution of relative energy with Wi, Re=0.5. Time is in a logarithmic scale.}  
     \label{energy_periodicu}
\end{figure*}

\begin{figure}
   \centering 
     \includegraphics[width=0.9\linewidth]{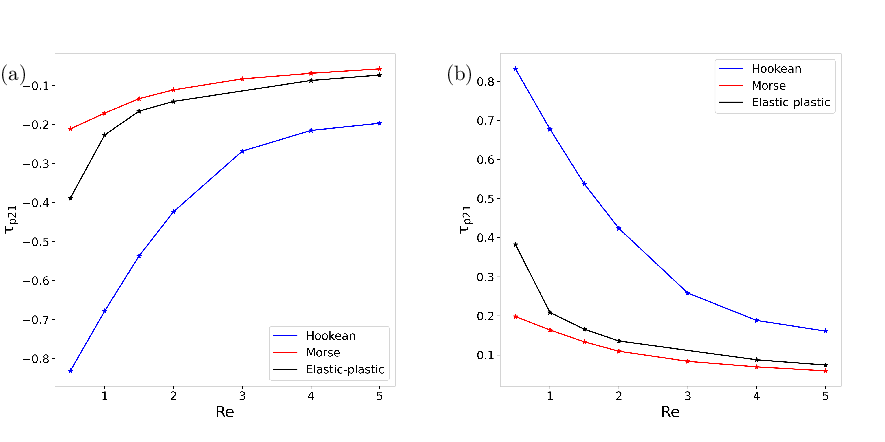}
   \caption{ The periodic $U$ case: The minimum (a) and maximum (b) value of shear stress $\tau_{p21}$ in one period as a function of Re on location $y=0.5$. }  
   \label{tau_Re_periodicu}
\end{figure}

\begin{figure*}
   \centering
    \includegraphics[width=0.9\linewidth]{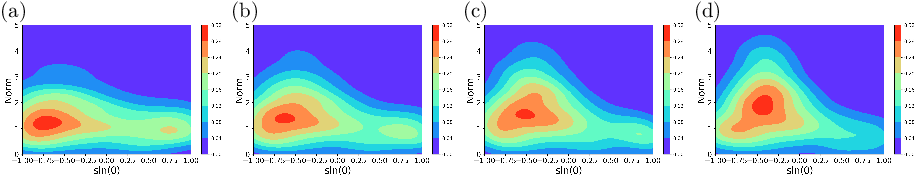}
   \caption{Hookean model in the periodic $U$ case: Contour plot of particle distribution as a function of $\sin \theta$ and norm at steady states when shear stresses take the minimum value of one period (from (a) to (d): Wi, Re=0.5, Wi, Re=1, Wi, Re=2 and Wi, Re=5).}
 \label{viscoplastic_KDE_periodicu}
\end{figure*}


Next, we check the energy evolution in this case. 
The time evolution of different types of energy for three models are presented in Figure \ref{energy_periodicu}.  
It clearly shows that the energies show periodic fluctuations. Moreover, 
for the Hookean model, the entropy energy and the potential energy almost oscillate around a constant as time evolves. This is expected since Figure \ref{hookperiodicKDE} shows no significant changes in the particle distribution for different times. 
%
For Morse and Elastic-plastic models, the input energy is converted into potential energy as in the case of constant U. And thus, the potential energy is increasing with periodic fluctuations. 
%
%
Meantime, the entropy energy is decreasing continuously with periodical fluctuations, which is similar to the case of constant U. It reveals that as time evolves, the particles are more and more ordered. 
This is consistent with the results in Figures \ref{plasticperiodicKDE2} and \ref{morseperiodicKDE2} that particles tend to be perpendicular to the moving plane when $t$ is large.

In summary, for Morse and Elastic-plastic models, both particle distribution and energy evolution indicate that the decrease in the magnitude of the stress $\tau_{p21}$ is mainly due to the following two reasons. One reason is the bond breaking: particles are elongated and bond breaking occurs as time evolves, causing the entropic force ($\nabla_{\bar{\qvec}_i}\Psi$) to become zero for the broken particles. 
The other reason is that many remaining unbroken particles tend to be perpendicular to the moving plane as time evolves.
As a result, $[\boldsymbol{\bar{q}}_i]_1$ approaches zero, and by the definition of shear stress, $\tau_{p21}$ tends to decrease.

\subsubsection{Rheology of the models with periodic $U$}

We analyze the stress $\tau_{p21}$ as a function of Re in the steady state. Since the time evolution of shear stress shows periodic behavior, 
Figure \ref{tau_Re_periodicu} presents the minimum and maximum values in one period at the steady state. It reveals that all three models show shear-thinning behavior, which is consistent with Figure \ref{periodic_shearstress_tplot}. The shear-thinning effect can also be explained by the contour plots of particle distributions. 
For illustration, only the contour plots for the Hookean model are presented in Figure \ref{viscoplastic_KDE_periodicu}.
Actually, the "pear" shaped plots presented in Figure \ref{viscoplastic_KDE_periodicu} reveal that with larger values of Re and Wi, more particles tend to align along the flow direction  (namely, $\sin \theta= 0$), resulting in $[\boldsymbol{\bar{q}}_i]_2$ approaching zero. Due to the definition of shear stress, this leads to fewer particles contributing to the stress $\tau_{p21}$, and hence the magnitude of $\tau_{p21}$ becomes smaller.

Figure \ref{tau_Re_periodicu} also shows that for Morse and Elastic-plastic models, values of stress $\tau_{p21}$ have a smaller magnitude (closer to 0) than the Hookean model as Re varies.
To explain this, we can refer to the contour plots of particle distribution with Wi, Re =0.5 in Figures \ref{morseperiodicKDE2} and \ref{plasticperiodicKDE2}. Compared to the contour plots of the Hookean model in Figure \ref{hookperiodicKDE2}, they illustrate that many particles are elongated significantly due to bond breaking, and the remaining particles with finite norm tend to be perpendicular to the moving plane when time $t$ is large. Thus, by the definition of shear stress, the values of $\tau_{p21}$ are closer to zero at the steady state.

 \vspace{0.5cm}
 
\section{Conclusion}

This work has investigated the micro-macro model for polymeric fluids, with various microscopic potential energies, including the classical Hookean potential, the newly proposed modified Morse potential, and Elastic-plastic potential. The study rederives the thermodynamically consistent micro-macro model by using the energy variational method. It comprehensively compares the microscopic configurations, macroscopic velocities, and shear stresses under shear flow for these three models, with a particular focus on the Morse and Elastic-plastic potentials, which incorporate bond-breaking occurrence at the microscopic scale. This exploration serves as a valuable complement to existing work in the field.

The numerical simulations are executed using a deterministic particle-FEM method, as originally proposed in Ref. \cite{bao2022},  which merges the deterministic particle method at the micro-scale with the finite element method at the macro-scale. Despite the deterministic particle method being a coarse-grained technique, it offers insights into the distribution of actual polymer chains. In this study, the configurations of microscopic polymer chains are inferred by examining the configurations of deterministic particles. The numerical findings shed light on the behaviors of polymer chains operating under three distinct microscopic potential energies. They also elucidate the corresponding behaviors of velocity and induced shear stresses of fluids at the macro-scale under shear flow conditions. Substantial differences between the Morse, Elastic-plastic models, and the classical Hookean model are explicated from two crucial angles: particle distribution and energy evolution. In cases of constant Couette flow, all three models exhibit an overshoot phenomenon in their velocity profiles. The curves of the Morse and Elastic-plastic models lie between those of the classical Hookean model and the Newtonian fluid, primarily due to bond breaking. This phenomenon introduces velocity fluctuations and shear stress variations, notably in the Elastic-plastic model. Furthermore, a reduction in shear stresses to zero is observed for the Morse and Elastic-plastic models, attributed to bond breaking and particle alignment with flow direction. The classical Hookean model exhibits shear-thinning behavior, due to increased polymer alignment along the flow direction at higher shear rates.

When the bottom plate undergoes periodic oscillations, all three models exhibit oscillatory behavior in velocity and shear stress evolutions. Additionally, the magnitude of polymer-induced stress for the Morse and Elastic-plastic models tend to decrease, whereas that for the Hookean model remains at constant. It is primarily
because some polymer chains elongate and become unbounded due to bond breaking. And the remaining bounded polymer chains align their directions perpendicular to the moving plane at various shear rates.
Furthermore, all three models demonstrate shear-thinning properties, as a larger proportion of polymer chains tend to align with the flow direction at higher shear rates. 

We need to note that this study assumes that it is uniform in the x-direction with a focus on 1D simple shear flow and assumes a dumbbell model for polymer chains. 
This can not describe the bending and folding of polymer chains and other complicated scenarios in 2D setting.  That is the limit of the present work. We will explore micro-macro models for multi-bead polymer chains and provide numerical results in 2D flows in future works.

\section*{Acknowledgments}
This work was partially supported by the 
National Natural Science Foundation of China no. 12071190, 12231004, 12201050, China Postdoctoral Science Foundation grant No. 2022M710425, and Natural Sciences and Engineering Research Council of Canada (NSERC).

\newpage

\bibliography{MM.bib}

\begin{thebibliography}{10}

\bibitem{Ammar2010}
A.~Ammar.
\newblock Lattice boltzmann method for polymer kinetic theory.
\newblock {\em J. Non-Newtonian Fluid Mech.}, 165:1082--1092, 2010.

\bibitem{ACR2006}
A.~Ammar, F.~Chinesta, and D.~Ryckelynck.
\newblock Deterministic particle approach of multi bead-spring polymer models.
\newblock {\em European J. Comput. Mech.}, 15(5):481--494, 2006.

\bibitem{viscoplastic}
N.~J. Balmforth, I.~A. Frigaard, G.~Ovarlez, and L.~Parns.
\newblock Yielding to stress: Recent developments in viscoplastic fluid
  mechanics.
\newblock {\em Annu. Rev. Fluid Mech.}, 46:121--146, 2014.

\bibitem{bao2021}
X.~Bao, R.~Chen, and H.~Zhang.
\newblock Constraint-preserving energy-stable scheme for the 2d simplified
  ericksen-leslie system.
\newblock {\em J. Comput. Math.}, 39(1):1--21, 2021.

\bibitem{bao2022}
X.~Bao, C.~Liu, and Y.~Wang.
\newblock On a deterministic particle-fem discretization to micro-macro models
  of dilute polymeric fluids.
\newblock {\em arXiv:2112.10970}, 2022.

\bibitem{becker2008}
R.~Becker, X.~Feng, and A.~Prohl.
\newblock Finite element approximations of the ericksen-leslie model for
  nematic liquid crystal flow.
\newblock {\em SIAM J. Numer. Anal.}, 46(4):1704--1731, 2008.

\bibitem{bird1987}
R.~B. Bird, C.~F. Curtiss, R.~C. Armstrong, and O.~Hassager.
\newblock {\em Dynamics of Polymeric Liquids, Kinetic Theory (Dynamics of
  Polymer Liquids, vol. 1 and 2)}.
\newblock Wiley-Interscience, New York, 1987.

\bibitem{bird1992transport}
R.~B. Bird and H.~C. {\"O}ttinger.
\newblock Transport properties of polymeric liquids.
\newblock {\em Annu. Rev. Phys. Chem.}, 43(1):371--406, 1992.

\bibitem{le2012micro}
C.~Le Bris and T.~Leli\ evre.
\newblock Micro-macro models for viscoelastic fluids: modelling, mathematics
  and numerics.
\newblock {\em Sci. China Math.}, 55(2):353--384, 2012.

\bibitem{Bris2012}
C.~Le Bris and T.~Leli\`{e}vre.
\newblock Micro-macro models for viscoelastic fluids: modelling, mathematics
  and numerics.
\newblock {\em Sci. China Math.}, 55(2):353--384, 2012.

\bibitem{Carrilloblob}
J.~A. Carrillo, K.~Craig, and F.~S. Patacchini.
\newblock A blob method for diffusion.
\newblock {\em Calc. Var.}, 58(2):53, 2019.

\bibitem{chenrui2015}
R.~Chen, G.~Ji, X.~Yang, and H.~Zhang.
\newblock Decoupled energy stable schemes for phase-field vesicle membrane
  model.
\newblock {\em J. Comput. Phys.}, 302:509--523, 2015.

\bibitem{Chorin1969}
A.~J. Chorin.
\newblock On the convergence of discrete approximations to the navier-stokes
  equations.
\newblock {\em Math. Comput.}, 23:341--353, 1969.

\bibitem{Doyle1998}
P.~S. Doyle, E.~S.G. Shaqfeh, G.~H. McKinley, and S.~H. Spiegelberg.
\newblock Relaxation of dilute polymer solutions following extensional flow.
\newblock {\em J. Non-Newtonian Fluid Mech.}, 76:79--110, 1998.

\bibitem{eisenberg2010energy}
B.~Eisenberg, Y.~Hyon, and C.~Liu.
\newblock Energy variational analysis of ions in water and channels: Field
  theory for primitive models of complex ionic fluids.
\newblock {\em J. Chem. Phys.}, 133(10):104104, 2010.

\bibitem{Filho2013}
R.~N.~Costa Filho, G.~Alencar, B.-S. Skagerstam, and J.~S.~Andrade Jr.
\newblock Morse potential derived from first principles.
\newblock {\em EPL}, 101(1):10009, 4 pp., 2013.

\bibitem{Glowinski2011}
R.~Glowinski and A.~Wachs.
\newblock On the numerical simulation of viscoplastic fluid flow.
\newblock {\em Handbook of Numerical Analysis}, 16:483--717, 2011.

\bibitem{Grmela1997I}
M.~Grmela and H.~C. \"{O}ttinger.
\newblock Dynamics and thermodynamics of complex fluids. i. development of a
  general formalism.
\newblock {\em Phys. Rev. E}, 56(6):6620--6632, 1997.

\bibitem{Guermond2006}
J.~L. Guermond, P.~Minev, and J.~Shen.
\newblock An overview of projection methods for incompressible flows.
\newblock {\em Comput. Methods Appl. Mech. Engrg.}, 195(44-47):6011--6045,
  2006.

\bibitem{Halin1998}
P.~Halin, G.~Lielens, R.~Keunings, and V.~Legat.
\newblock The lagrangian particle method for macroscopic and micro-macro
  viscoelastic flow computations.
\newblock {\em J. Non-Newtonian Fluid Mech.}, 79:387--403, 1998.

\bibitem{Hulsen1997}
M.~A. Hulsen, A.~P.~G. van Heel, and B.~H. A.~A. van~den Brule.
\newblock Simulation of viscoelastic flows using brownian configuration fields.
\newblock {\em J. Non-Newtonian Fluid Mech.}, 70:79--101, 1997.

\bibitem{Keunings1997}
R.~Keunings.
\newblock On the peterlin approximation for finitely extensible dumbbells.
\newblock {\em J. Non-Newtonian Fluid Mech.}, 68:85--100, 1997.

\bibitem{Laso1993}
M.~Laso and H.~C. \"{O}ttinger.
\newblock Calculation of viscoelastic flow using molecular models: The
  connffessit approach.
\newblock {\em J. Non-Newtonian Fluid Mech.}, 47:1--20, 1993.

\bibitem{Lielens1998}
G.~Lielens, P.~Halin, I.~Jaumain, R.~Keunings, and V.~Legat.
\newblock New closure approximations for the kinetic theory of finitely
  extensible dumbbells.
\newblock {\em J. Non-Newtonian Fluid Mech.}, 76:249--279, 1998.

\bibitem{lin-liu-zhang}
F.~Lin, C.~Liu, and P.~Zhang.
\newblock On a micro-macro model for polymeric fluids near equilibrium.
\newblock {\em Comm. Pure Appl. Math.}, 60(6):838--866, 2007.

\bibitem{wang2020jcp}
C.~Liu and Y.~Wang.
\newblock On {L}agrangian schemes for porous medium type generalized diffusion
  equations: A discrete energetic variational approach.
\newblock {\em J. Comput. Phys.}, 417:109566, 2020.

\bibitem{needleman2017active}
D.~Needleman and Z.~Dogic.
\newblock Active matter at the interface between materials science and cell
  biology.
\newblock {\em Nat. Rev. Mater.}, 2(9):1--14, 2017.

\bibitem{nguyen2017single}
T-H. Nguyen.
\newblock Single-molecule force spectroscopy applied to heparin-induced
  thrombocytopenia.
\newblock {\em J. Mol. Recognit.}, 30(3):e2585, 2017.

\bibitem{ottinger1996}
H.~C. \"{O}ttinger.
\newblock {\em Stochastic Processes in Polymeric Fluids, Tools and Examples for
  Developing Simulation Algorithms}.
\newblock Springer-Verlag, Berlin, 1996.

\bibitem{Grmela1997II}
H.~C. \"{O}ttinger and M.~Grmela.
\newblock Dynamics and thermodynamics of complex fluids. ii. illustrations of a
  general formalism.
\newblock {\em Phys. Rev. E}, 56(6):6633--6655, 1997.

\bibitem{owens2002}
R.~G. Owens and T.~N. Phillips.
\newblock {\em Computational Rheology}.
\newblock Imperial College Press, London, 2002.

\bibitem{Papanastasiou1997}
T.~C. Papanastasiou and A.~G. Boudouvis.
\newblock Flows of viscoplastic materials: Models and computations.
\newblock {\em Comput. Struct.}, 64(1-4):677--694, 1997.

\bibitem{shen2020energy}
L.~Shen, H.~Huang, P.~Lin, Z.~Song, and S.~Xu.
\newblock An energy stable c0 finite element scheme for a quasi-incompressible
  phase-field model of moving contact line with variable density.
\newblock {\em J. Comput. Phys.}, 405:109179, 2020.

\bibitem{shen2022energy}
L.~Shen, Z.~Xu, P.~Lin, H.~Huang, and S.~Xu.
\newblock An energy stable c\^{0} finite element scheme for a phase-field model
  of vesicle motion and deformation.
\newblock {\em SIAM J. Sci. Comput.}, 44(1):B122--B145, 2022.

\bibitem{Sizaire1999}
R.~Sizaire, G.~Lielens, I.~Jaumain, R.~Keunings, and V.~Legat.
\newblock On the hysteretic behaviour of dilute polymer solutions in relaxation
  following extensional flow.
\newblock {\em J. Non-Newtonian Fluid Mech.}, 82:233--253, 1999.

\bibitem{tm11}
E.~B. Tadmor and R.~E. Miller.
\newblock {\em Modeling materials. Continuum, Atomistic and Multiscale
  Techniques}.
\newblock Cambridge University Press, Cambridge, 2011.

\bibitem{Theodorakopoulos2000}
N.~Theodorakopoulos, T.~Dauxois, and M.~Peyrard.
\newblock Order of the phase transition in models of dna thermal denaturation.
\newblock {\em Phys. Rev. Lett.}, 85(1):6--9, 2000.

\bibitem{wangEVI}
Y.~Wang, J.~Chen, L.~Kang, and C.~Liu.
\newblock Particle-based energetic variational inference.
\newblock {\em Stat. Comput.}, 31(3):Paper No. 34, 17pp., 2021.

\bibitem{wang2021two}
Y.~Wang, T-F. Zhang, and C.~Liu.
\newblock A two species micro--macro model of wormlike micellar solutions and
  its maximum entropy closure approximations: An energetic variational
  approach.
\newblock {\em Journal of Non-Newtonian Fluid Mechanics}, 293:104559, 2021.

\bibitem{xu2019three}
S.~Xu, M.~Alber, and Z.~Xu.
\newblock Three-phase model of visco-elastic incompressible fluid flow and its
  computational implementation.
\newblock {\em Commun. Comput. Phys.}, 25(2):586, 2019.

\bibitem{xu2014energetic}
S.~Xu, P.~Sheng, and C.~Liu.
\newblock An energetic variational approach for ion transport.
\newblock {\em Commun. Math. Sci.}, 12(4):779--789, 2014.

\bibitem{xu2017model}
S.~Xu, Z.~Xu, O.~V. Kim, R.~I. Litvinov, J.~W. Weisel, and M.~Alber.
\newblock Model predictions of deformation, embolization and permeability of
  partially obstructive blood clots under variable shear flow.
\newblock {\em J. R. Soc. Interface}, 14(136):20170441, 2017.

\bibitem{XuSPH2014}
X.~Xu, J.~Ouyang, W.~Li, and Q.~Liu.
\newblock Sph simulations of 2d transient viscoelastic flows using brownian
  configuration fields.
\newblock {\em J. Non-Newtonian Fluid Mech.}, 208-209:59--71, 2014.

\bibitem{xu2016multi}
X.~Xu and P.~Yu.
\newblock A multiscale sph method for simulating transient viscoelastic flows
  using bead-spring chain model.
\newblock {\em J. Non-Newtonian Fluid Mech.}, 229:27--42, 2016.

\end{thebibliography}
\end{document}